\documentclass[aps,pre,preprint,a4paper,onecolumn]{revtex4}

\usepackage[]{amsmath}
\usepackage{amssymb}
\usepackage[dvips]{graphicx}
\usepackage{color}
\usepackage{tabularx}
\usepackage{mathtools}
\usepackage{algpseudocode}
\usepackage{enumitem}
\usepackage{multirow}
\usepackage{epstopdf}  
\usepackage{hyperref}


\newcommand{\vect}[1]{\textbf{\textit{#1}}}

\newcommand{\inp}{{\mathrm{in}}}
\newcommand{\out}{{\mathrm{out}}}

\newcommand{\myphi}{\varphi}

\newcommand{\mn}{\mathcal {N}}
\newcommand{\ml}{\mathcal {L}}

\begin{document}

\title{
Deep Potential Molecular Dynamics: 
a scalable model with the accuracy of quantum mechanics
}
\author{Linfeng Zhang}
\affiliation{Program in Applied and Computational Mathematics, 
Princeton University, Princeton, NJ 08544, USA}
\author{Jiequn Han}
\affiliation{Program in Applied and Computational Mathematics, 
Princeton University, Princeton, NJ 08544, USA}
\author{Han Wang}
\email{wang_han@iapcm.ac.cn}
\affiliation{Institute of Applied Physics and Computational Mathematics,
Fenghao East Road 2, Beijing 100094, P.R.~China}
\affiliation{CAEP Software Center for High Performance Numerical
Simulation, Huayuan Road 6, Beijing 100088, P.R.~China}
\author{Roberto Car}
\affiliation{Department of Chemistry,  
Department of Physics, 
Program in Applied and Computational Mathematics, 
Princeton Institute for the Science and Technology of Materials,  
Princeton University, Princeton, NJ 08544, USA}
\author{Weinan E}
\email{weinan@math.princeton.edu}
\affiliation{Department of Mathematics and Program 
in Applied and Computational Mathematics, 
Princeton University, Princeton, NJ 08544, USA}
\affiliation{Center for Data Science, 
Beijing International Center for Mathematical Research,  Peking University,
Beijing Institute of Big Data Research, 
Beijing, 100871, P.R.~China}

\begin{abstract}
We introduce a scheme for molecular simulations,
the Deep Potential Molecular Dynamics (DeePMD) method, 
based on a many-body potential and interatomic forces generated 
by a carefully crafted deep neural network trained with \textit{ab initio} data. 
The neural network model preserves all the natural symmetries in the problem.
It is ``first principle-based'' in the sense that there are
no \textit{ad hoc} components aside from the network model.
We show that the proposed scheme
provides an efficient and accurate protocol in a variety of systems,
including bulk materials and molecules.
In all these cases, 
DeePMD gives results that are essentially indistinguishable 
from the original data, at a cost that scales linearly with system size.
\end{abstract}

\maketitle
Molecular dynamics (MD) is used in many disciplines, 
including physics, chemistry, biology, and materials science, 
but its accuracy depends on the model for the atomic interactions. 
\textit{Ab initio} molecular dynamics (AIMD) \cite{car1985unified,marx2009ab}
has the accuracy of density functional theory (DFT) ~\cite{kohn1965self}, 
but its computational cost limits typical applications to 
hundreds of atoms and time scales of $\sim$100 $ps$.
Applications requiring larger cells and longer simulations
are currently accessible only with empirical force fields (FFs)
\cite{vanommeslaeghe2010charmm, jorgensen1996development,wang2004development},
but the accuracy and transferability of these models is often in question.

Developing FFs is challenging due to the many-body character of the potential energy. 
Expansions in 2- and 3-body interactions may capture the physics \cite{stillinger1985SW} but are strictly valid only for weakly interacting systems. 
A large class of potentials, including the embedded atom method (EAM) \cite{daw1984EAM}, the bond order potentials \cite{brenner2002BOP},
and the reactive FFs \cite{van2001rff}, share the physically motivated idea that the strength of a bond depends on the local environment,
but the functional form of this dependence can only be given with crude approximations.   

Machine learning (ML) methodologies are changing this state of affairs 
\cite{thompson2015SNAP,huan2017AGNI,behler2007generalized,behler2016perspective,
morawietz2016van,bartok2010gaussian,rupp2012fast,
schutt2017quantum,chmiela2017machine,smith2017ani}. 
When trained on large datasets of atomic configurations and corresponding potential energies and forces, 
ML models can reproduce the original data accurately. 
In training these models, the atomic coordinates cannot be used as they appear in MD trajectories because their format does not preserve 
the translational, rotational, and permutational symmetry of the system. 
Different ML models address this issue in different ways.
Two successful schemes are 
the Behler-Parrinello neural network (BPNN) \cite{behler2007generalized} 
and the gradient-domain machine learning (GDML) \cite{chmiela2017machine}. 
In BPNN symmetry is preserved by mapping the coordinates onto a large set of two- and three-body symmetry functions,
which are, however, largely {\it ad hoc}.
Fixing the symmetry functions may become painstaking in systems with many atomic species. 
In GDML the same goal is achieved by mapping the coordinates onto the eigenvalues of the Coulomb matrix, 
whose elements are the inverse distances between all distinct pairs of atoms.
It is not straightforward how to use the Coulomb matrix in extended periodic systems.
So far GDML has only been used for relatively small molecules.

In this letter we introduce an NN scheme for MD simulations, 
called Deep Potential Molecular Dynamics (DeePMD), 
which overcomes the limitations associated to
auxiliary quantities like the symmetry functions or the Coulomb matrix. 
In our scheme a local reference frame and a local environment is assigned to each atom.
Each environment contains a finite number of atoms, whose local coordinates are
arranged in a symmetry preserving way following the prescription of the Deep Potential method \cite{han2017deep},
an approach that was devised to train an NN with the potential energy only.
With typical AIMD datasets this is insufficient to reproduce the trajectories.
DeePMD overcomes this limitation. In addition, the learning process in DeePMD improves significantly 
over the Deep Potential method thanks to the introduction of a flexible family of loss functions.   
The NN potential constructed in this way reproduces accurately the AIMD trajectories, 
both classical and quantum (path integral), in extended and finite systems, 
at a cost that scales linearly with system size and is always several orders of magnitude lower than that of equivalent AIMD simulations.

In DeePMD the potential energy of each atomic configuration is a sum of ``atomic energies", $E = \sum_i E_i$,
where $E_i$ is determined by the local environment of atom $i$ within a cutoff radius $R_c$ and can be seen as a realization of the embedded atom concept. 
The environmental dependence of $E_i$, which embodies the many-body character of the interactions, is complex and nonlinear. 
The NN is able to capture the analytical dependence of $E_i$ on the coordinates of the atoms in the environment 
in terms of the composition of the sequence of mappings associated to the individual hidden layers. 
The additive form of $E$ naturally preserves the extensive character of the potential energy. 
Due to the analyticity of the ``atomic energies'' DeePMD is, in principle, a conservative model.

\begin{figure}
  \centering
  \includegraphics[width=0.38\textwidth]{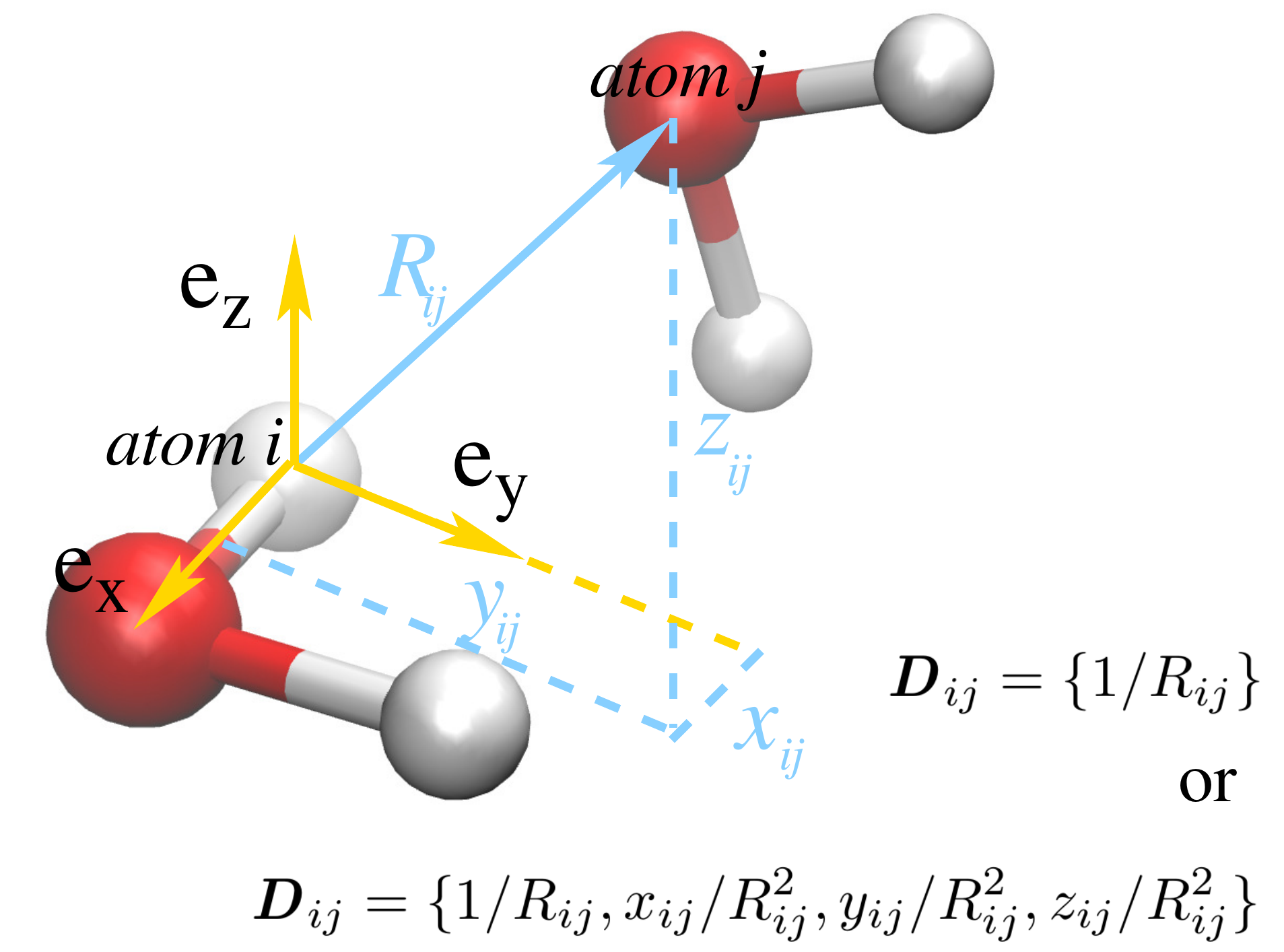}
  \caption{(color online). Schematic plot of the neural network input
  for the environment of atom $i$, taking water as an example.
    Atom $j$ is a generic neighbor of atom $i$.
    $(\vect e_x, \vect e_y, \vect e_z)$ is the local frame of atom $i$.
    $\vect e_x$ is along the O-H bond. $\vect e_z$ is
    perpendicular to the plane of the water molecule.
    $\vect e_y$ is the cross product of $\vect e_z$ and $\vect e_x$.
    $(x_{ij}, y_{ij}, z_{ij})$ are the Cartesian components of the vector
    $\vect R_{ij}$ in this local frame.
    $R_{ij}$ is the length of $\vect R_{ij}$.
    The neural network input $\vect D_{ij}$
    may either contain the full radial 
    and angular information of atom $j$,
    i.e.,~$\vect D_{ij} = \{1/ R_{ij}, x_{ij}/R^2_{ij}, 
    y_{ij}/R^2_{ij}, z_{ij}/R^2_{ij}\}$, 
    or only the radial information, 
    i.e.,~$\vect D_{ij} = \{1/ R_{ij}\}$.
    We first sort the neighbors of atom $i$ 
    according to their chemical species, 
    e.g.~oxygens first then hydrogens.
    Within each species we sort the atoms according to 
    their inverse distances to atom $i$, i.e.,~$1/R_{ij}$.
    We use $\{\vect D_{ij}\}$ to denote the sorted input data for atom $i$.
  }
  \label{fig:local-frame}
\end
{figure}

$E_i$ is constructed in two steps. 
First, a local coordinate frame is set up
for every atom and its neighbors inside $R_c$
\footnote{Some flexibility can be used in the definition 
of the local frame of atom $i$. 
Usually we define it in terms of the two atoms closest to $i$, 
independently of their species.
Exceptions to this rule are discussed in the SM.}.
This allows us to preserve the translational, rotational, and 
permutational symmetries of the environment, as shown in Fig.~\ref{fig:local-frame},
which illustrates the format adopted for the local coordinate information $(\{\vect D_{ij}\})$.
The $1/R_{ij}$ factor present in $\vect D_{ij}$ reduces the weight of the particles 
that are more distant from atom $i$.

\begin{figure}
  \centering
  \includegraphics[width=0.47\textwidth]{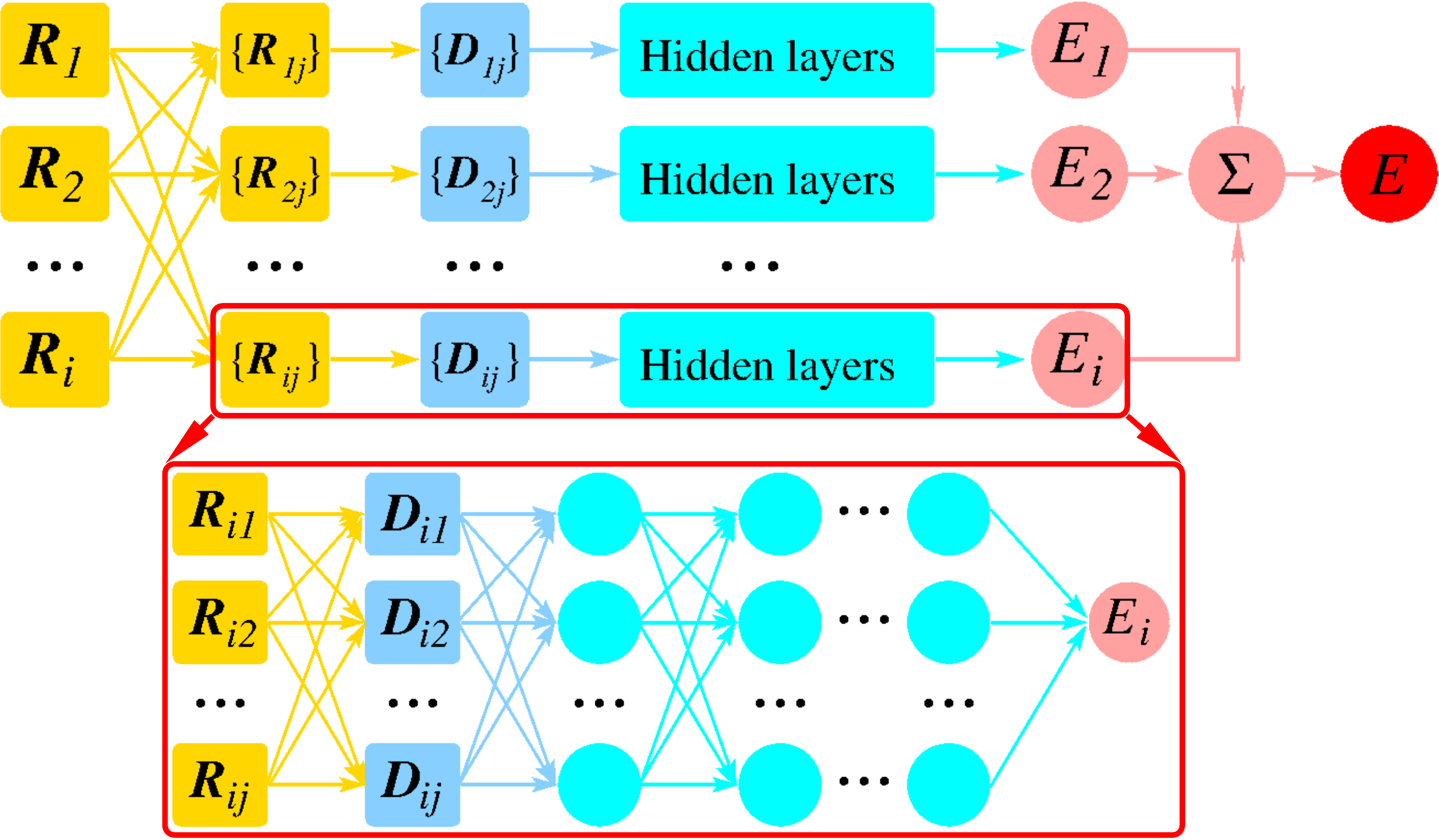}
  \caption{(color online). Schematic plot of the DeePMD model.
    The frame in the box is the zoom-in of a DNN.
    The relative positions of all neighbors w.r.t.~atom $i$, 
    i.e.,~$\{\vect R_{ij}\}$,
    is first converted to $\{\vect D_{ij}\}$,
    then passed to
    the hidden layers to compute $E_i$.
  }
  \label{fig:deep-pot}
\end{figure}

Next, $\{\vect D_{ij}\}$ serves as input of 
a deep neural network (DNN)~\cite{goodfellow2016dl}, 
which returns $E_i$ in output (Fig.~\ref{fig:deep-pot}). 
The DNN is a feed forward network, 
in which data flow from the input layer
to the output layer ($E_i$),
through multiple hidden layers consisting of several nodes 
that input the data $d^\inp_l$ from the previous layer
and output the data $d^\out_k$ to the next layer.
A linear transformation is applied to the input data, i.e.,
$  \tilde{d}_k = \sum_{l} w_{kl} d^\inp_l + b_k $,
followed by action of a non-linear function $\myphi$ on $\tilde{d}_k$,
i.e., $d^\out_k = \myphi \big( \tilde{d}_k  \big)$. 
In the final step from the last hidden layer to $E_{i}$, 
only the linear transformation is applied. 
The composition of the linear and nonlinear transformations introduced above provides the analytical representation of
$E_i$ in terms of the local coordinates.
The technical details of this construction are discussed in the supplementary materials (SM).
In our applications, we adopt the hyperbolic tangent for $\myphi$
and use 5 hidden layers with decreasing number of nodes per layer,
i.e., 240, 120, 60, 30, and 10 nodes, respectively, 
from the innermost to the outermost layer.
It is known empirically that the hidden layers greatly enhance the capability of neural networks 
to fit complex and highly nonlinear functional dependences \cite{bengio2007greedy,krizhevsky2012imagenet}.
In our case, only by including a few hidden layers could DeePMD reproduce the trajectories with sufficient accuracy.

We use the Adam method \cite{Kingma2015adam} to optimize the parameters $w_{kl}$ and $b_k$ of each layer
with the family of loss functions
\begin{align}\label{eqn:loss}
  \hspace{-.4cm}
  L(p_\epsilon, p_f, p_\xi) = {p_\epsilon} \Delta \epsilon^2 
  + \frac{p_f}{3N} \sum_i \vert \Delta\vect F_i\vert^2 
  + \frac{p_\xi}{9} ||\Delta \xi||^2.
\end{align}
Here $\Delta$ denotes the difference between the DeePMD prediction and the training data, 
$N$ is the number of atoms, $\epsilon$ is the energy per atom, 
$ \vect F_i$ is the force on atom $i$, 
and $ \xi$ is the virial tensor $\Xi=-\frac{1}{2}\sum_i \vect R_i\otimes\vect F_i$ divided by $N$.
In Eq.~\eqref{eqn:loss}, $p_\epsilon$, $p_f$, and $p_\xi$ 
are tunable prefactors. 
When virial information is missing from the data, we set $p_\xi=0$. 
In order to minimize the loss function in Eq.~\eqref{eqn:loss} in a well balanced way, 
we vary the magnitude of the prefactors during training.
We progressively increase $p_\epsilon$ and $p_\xi$ and decrease $p_f$, 
so that the force term dominates at the beginning, 
while energy and virial terms become important at the end. 
We find that this strategy is very effective and reduces the total training time to a few core hours in all the test cases. 

To test the method, 
we have applied DeePMD to extended and finite systems.
As representative extended systems, we consider
(a) liquid water at $P$ = 1 bar and $T$ = 300 K, at the PI-AIMD level,
(b) ice Ih at $P$ = 1 bar and $T$ = 273 K, at the PI-AIMD level,
(c) ice Ih at $P$ = 1 bar and $T$ = 330 K, at the classical AIMD level, and
(d) ice Ih at $P$ = 2.13 kbar and $T$ = 238 K, 
which is the experimental triple point for ice I, II, and III,
at the classical AIMD level.
The variable periodic simulation cell contains 
64 H${}_2$O molecules in the case of liquid water, 
and 96 H${}_2$O molecules in the case of ices.
We adopt $R_c$ = 6.0~\AA~and use the full radial and angular information
for the 16 oxygens and the 32 hydrogens closest to the atom at the origin, 
while retaining only radial information for all the other atoms within $R_c$.
All the ice simulations include proton disorder.
Deuterons replace protons in the simulations (c) and (d).
The PBE0+TS \cite{Carlo1999PBE0,TS2009TS} functional is adopted in all cases.
As representative finite systems we consider benzene, uracil, napthalene, aspirin, salicylic acid, 
malonaldehyde, ethanol, and toluene, for which classical AIMD trajectories with the PBE+TS functional 
\cite{Perdew1996PBE,TS2009TS}  
are available~\footnote{http://quantum-machine.org/}.
In these systems, we set $R_c$ large enough to include all the atoms,
and use the full radial and angular information in each local frame.

We discuss the performance of DeePMD according to four criteria:
($i$) generality of the model; 
($ii$) accuracy of the energy, forces, and virial tensor;
($iii$) faithfulness of the trajectories; 
($iv$) scalability and computational cost.
We refer to the SM for full details 
on the DeePMD implementation and the training datasets.

$Generality$.
Bulk and molecular systems exhibit different levels of complexity.
The liquid water samples include quantum fluctuations.
The organic molecules differ in composition and size,
and the corresponding datasets 
include large numbers of conformations. 
Yet DeePMD produces satisfactory results in all cases, 
using the same methodology, network structure, and optimization scheme.
The excellent performance of DeePMD in systems so diverse suggests that the method should 
be applicable to harder systems such as biological molecules, alloys, and liquid mixtures.

$Accuracy$.
We quantify the accuracy of energy, 
forces, and virial predictions in terms of the root mean square error
(RMSE) in the case of water and ices (Tab.~\ref{tab:error-water}), 
and in terms of the mean absolute error (MAE) 
in the case of the organic molecules (Tab.~\ref{tab:error-mol}).
No virial information was used for the latter.
In the water case,
the RMSE of the forces is comparable to the accuracy of the
minimization procedure in the original AIMD simulations, 
in which the allowed error in the forces was less than $10^{-3}$~a.u..
In the case of the molecules, the predicted energy and forces 
are generally slightly better than the GDML benchmark.

\begin{table}
  \centering
  \caption{The RMSE of the DeePMD prediction 
  for water and ices in terms of the energy, 
the forces, and/or the virial. 
The RMSEs of the energy and the virial 
are normalized by the number of molecules in the system. 
}
  \label{tab:error-water}
  \begin{tabular*}{0.48\textwidth}{@{\extracolsep{\fill}}
    cccc}\hline\hline
    System  &{Energy [meV]} &{Force [meV/\AA]} & {Virial [meV]}  \\\hline
    liquid water    & 1.0        & 40.4    & 2.0\\
    ice Ih (b)      & 0.7        & 43.3    & 1.5  \\
    ice Ih (c)      & 0.7        & 26.8    & -  \\
    Ice Ih (d)      & 0.8        & 25.4    & -  \\\hline\hline
  \end{tabular*}
\end{table}

\begin{table}
  \centering
  \caption{The MAE of the DeePMD prediction for organic molecules 
  in terms of the energy and the forces. The numbers in parentheses 
    are the GDML results~\cite{chmiela2017machine}.
  }
  \label{tab:error-mol}
  \begin{tabular*}{0.48\textwidth}{@{\extracolsep{\fill}}
    lrlrl}\hline\hline
     Molecule    & \multicolumn{2}{c}{Energy [meV]}      &
     \multicolumn{2}{c}{Force [meV/\AA]}  \\\hline
    Benzene        & 2.8      &  (3.0)       &  7.6          & (10.0)\\
    Uracil         & 3.7      &  (4.0)       & 9.8           & (10.4)\\
    Naphthalene    & 4.1      &  (5.2)       &  7.1          & (10.0)\\
    Aspirin        & 8.7      & (11.7)       & 19.1          & (42.9)\\
    Salicylic acid & 4.6      & (5.2)        & 10.9          & (12.1)\\
    Malonaldehyde  & 4.0      & (6.9)        & 12.7          & (34.7)\\
    Ethanol        & 2.4      & (6.5)        &  8.3          & (34.3)\\
    Toluene        & 3.7      & (5.2)        &  8.5          & (18.6) \\
    \hline\hline
  \end{tabular*}
\end{table}

\begin{figure}
  \centering
  \includegraphics [width=0.47\textwidth] {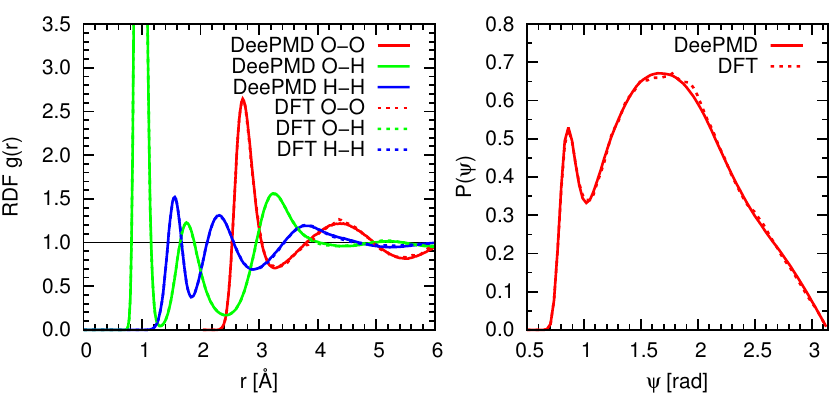}
  \caption{Correlation functions of liquid water from DeePMD and PI-AIMD. Left: RDFs. Right: the O-O-O ADF within a cutoff radius of 3.7 \AA.}
  \label{fig:comp-rdf}
\end{figure}

\begin{figure}
  \centering
  \includegraphics[width=0.47\textwidth]{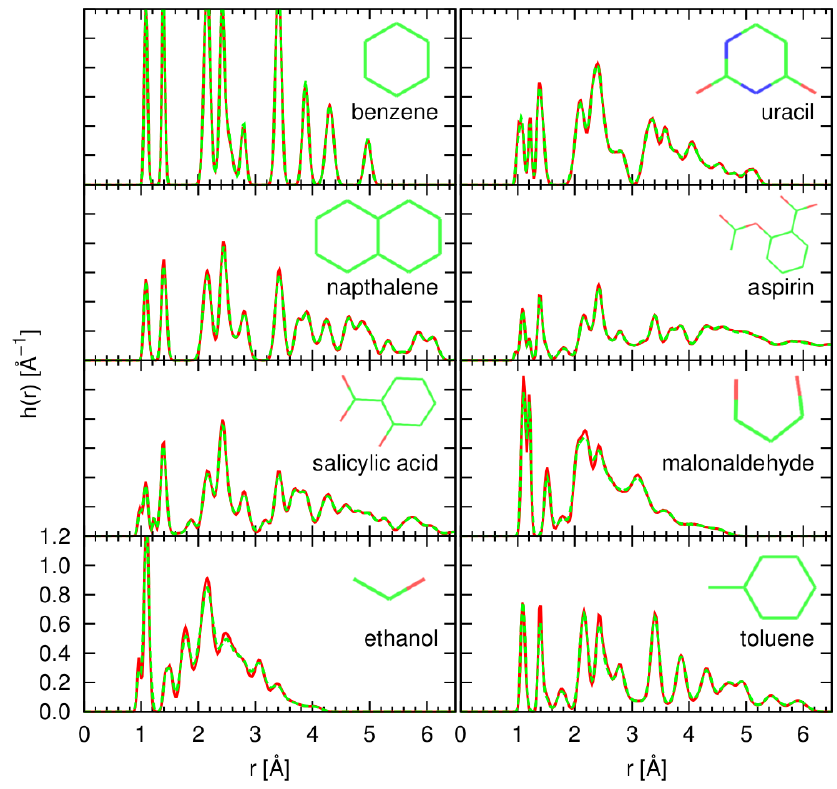}
  \caption{Interatomic distance
  distributions of the organic molecules.
    The solid lines denote the DeePMD results. 
    The dashed lines denote the AIMD results.
  }\label{fig:comp-idd}
\end{figure}

$MD$ $trajectories$.
In the case of water and ices, 
we perform path-integral/classical DeePMD simulations 
at the thermodynamic conditions of the original models, 
using the i-PI software \cite{Ceriotti2014iPI},
but with much longer simulation time (300~$ps$).
The average energy $\bar{E}$,
density $\bar{\rho}$, radial distribution functions (RDFs),
and a representative angular distribution function (ADF), i.e., a 3-body correlation function,
are reproduced with high accuracy. 
The results are summarized in Tab.~\ref{tab:equilibrium}.
The RDFs and ADF of the quantum trajectories of water are shown in Fig.~\ref{fig:comp-rdf}.
The RDFs of ice are reported in the SM. 
A higher-order correlation function, 
the probability distribution function of the O-O bond orientation order parameter $Q_6$, is additionally reported in the SM
and shows excellent agreement between DeePMD and AIMD trajectories.
In the case of the molecules, we perform DeePMD at the same 
temperature of the original data, 
using a Langevin thermostat with a damping time $\tau$ = 0.1 ps. 
The corresponding distributions of interatomic distances
are very close to the original data (Fig.~\ref{fig:comp-idd}). 

\begin{table}
  \centering
  \caption{The equilibrium energy and density, $\bar{E}$ and $\bar{\rho}$, 
  of water and ices, with DeePMD and AIMD. 
  The numbers in square brackets are the AIMD results.
  The numbers in parentheses are statistical uncertainties in the last one or two digits.
  The training AIMD trajectories for the ices are shorter and more correlated than in the water case.}
  \label{tab:equilibrium}
  \begin{tabular*}{0.48\textwidth}{@{\extracolsep{\fill}}
    lrlrl}\hline\hline
    System    & \multicolumn{2}{c}{$\bar{E}$[eV/H${}_2$O]} & 
    \multicolumn{2}{c}{$\bar{\rho}$[g/m${}^3$]}  \\\hline
    liquid water   & -467.678(2)   &[-467.679(6)]&  1.013(5)   &  [1.013(20)]  \\
    ice Ih (b)     & -467.750(1)   &[-467.747(4)]&  0.967(1)  &  [0.966(6)] \\
    ice Ih (c)     & -468.0478(3)   &[-468.0557(16)]&  0.950(1)  &  [0.949(2)] \\
    ice Ih (d)     & -468.0942(2)   &[-468.1026(9)]&  0.986(1)  &  [0.985(2)] \\
\hline\hline
\end{tabular*}
\end{table}

\begin{figure}
  \centering
  \includegraphics[width=0.4\textwidth]{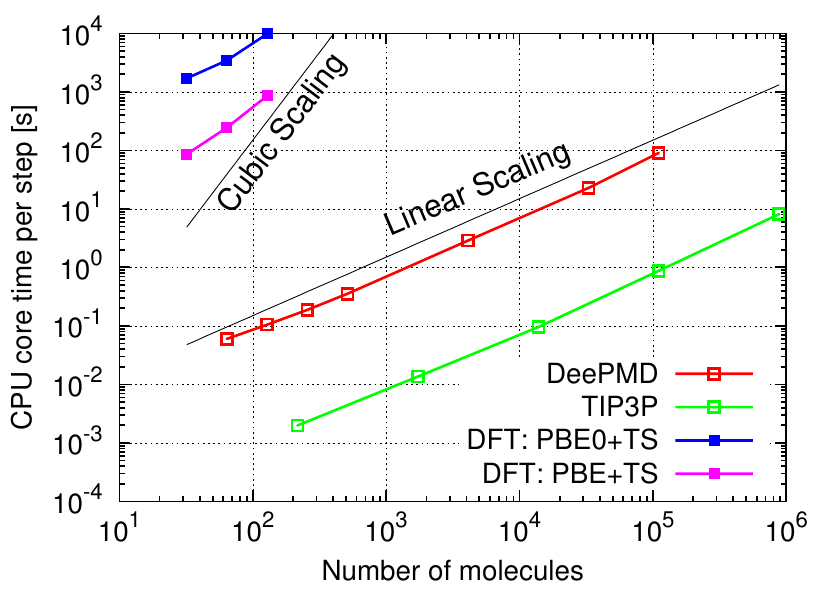}
  \caption{Computational cost of MD step $vs.$ system size, with DeePMD, 
  TIP3P, PBE+TS and, PBE0+TS.
  All simulations are performed on a Nersc Cori supercomputer with
  the Intel Xeon CPU E5-2698 v3.
  The TIP3P simulations use the Gromacs codes 
  (version 4.6.7) \cite{pronk2013gromacs}.
  The PBE+TS and PBE0+TS simulations use the 
  Quantum Espresso codes \cite{QE-2017}.}
  \label{fig:scaling}
\end{figure}

$Scalability$ $and$ $computational$ $cost$.
All the physical quantities in DeePMD are sums of local contributions. 
Thus, after training on a relatively small system, 
DeePMD can be directly applied to much larger systems.
The computational cost of DeePMD scales linearly with the number of atoms. 
Moreover, 
DeePMD can be easily parallelized due to its local decomposition
and the near-neighbor dependence of its ``atomic energies''.
In Fig.~\ref{fig:scaling}, 
we compare the cost of DeePMD fixed-cell simulations ($NVT$)
of liquid water with that of equivalent simulations with AIMD and the empirical FF
TIP3P~\cite{Jorgensen1983TIP3P} in units of CPU core seconds/step/molecule.

While in principle the environmental dependence of $E_i$ is analytical, 
in our implementation discontinuities are present in the forces, due to adoption of a sharp cutoff radius,
limitation of angular information to a fixed number of atoms, 
and abrupt changes in the atomic lists due to sorting. 
These discontinuities are similar in magnitude to those present in the AIMD forces due to finite numerical accuracy in
the enforcement of the Born-Oppenheimer condition.
In both cases, the discontinuities are much smaller than thermal fluctuations and perfect canonical evolution is achieved by coupling the systems to a thermostat.
We further note that long-range Coulomb interactions are not treated explicitly in the current implementation, 
although implicitly present in the training data. 
Explicit treatment of Coulombic effects may be necessary in some applications and deserves further study.  

In conclusion, DeePMD realizes a paradigm for molecular simulation, 
wherein accurate quantum mechanical data are faithfully parametrized by machine learning algorithms, 
which make possible simulations of DFT based AIMD quality on much larger systems 
and for much longer time than with direct AIMD. 
While substantially more predictive than empirical FFs, DFT is not chemically accurate 
\footnote{Conventionally, chemical accuracy corresponds to an error of 1 kcal/mol in the energy.}. 
In principle DeePMD could be trained with chemically accurate data from high-level quantum chemistry \cite{Watts1993CCSDT} and/or quantum Monte Carlo \cite{ceperley1986qmc}, 
but so far this has been prevented by the large computational cost of these calculations.

DeePMD should also very useful to coarse grain the atomic degrees of freedom,
for example, by generating an NN model for a reduced set of degrees of freedom while using the full set of degrees of freedom for training. 
The above considerations suggest that DeePMD should enhance considerably the
realm of AIMD applications by successfully addressing
the dilemma of accuracy versus efficiency that has confronted the molecular simulation community for a long time.

\begin{acknowledgements}
The authors acknowledge H.-Y. Ko and B. Santra 
for sharing the AIMD data on water and ice.
The work of J. Han and W. E is supported 
in part by Major Program of NNSFC under grant 91130005, 
ONR grant N00014-13-1-0338, DOE grants DE-SC0008626 and DE-SC0009248, and NSFC grant U1430237. 
The work of R. Car is supported in part by DOE-SciDAC grant DE-SC0008626.
The work of H. Wang is supported by the National Science Foundation 
of China under Grants 11501039 and 91530322, 
the National Key Research and Development Program of China 
under Grants 2016YFB0201200 and 2016YFB0201203, 
and the Science Challenge Project No. JCKY2016212A502.
\end{acknowledgements}


\begin{thebibliography}{34}%
\makeatletter
\providecommand \@ifxundefined [1]{%
 \@ifx{#1\undefined}
}%
\providecommand \@ifnum [1]{%
 \ifnum #1\expandafter \@firstoftwo
 \else \expandafter \@secondoftwo
 \fi
}%
\providecommand \@ifx [1]{%
 \ifx #1\expandafter \@firstoftwo
 \else \expandafter \@secondoftwo
 \fi
}%
\providecommand \natexlab [1]{#1}%
\providecommand \enquote  [1]{``#1''}%
\providecommand \bibnamefont  [1]{#1}%
\providecommand \bibfnamefont [1]{#1}%
\providecommand \citenamefont [1]{#1}%
\providecommand \href@noop [0]{\@secondoftwo}%
\providecommand \href [0]{\begingroup \@sanitize@url \@href}%
\providecommand \@href[1]{\@@startlink{#1}\@@href}%
\providecommand \@@href[1]{\endgroup#1\@@endlink}%
\providecommand \@sanitize@url [0]{\catcode `\\12\catcode `\$12\catcode
  `\&12\catcode `\#12\catcode `\^12\catcode `\_12\catcode `\%12\relax}%
\providecommand \@@startlink[1]{}%
\providecommand \@@endlink[0]{}%
\providecommand \url  [0]{\begingroup\@sanitize@url \@url }%
\providecommand \@url [1]{\endgroup\@href {#1}{\urlprefix }}%
\providecommand \urlprefix  [0]{URL }%
\providecommand \Eprint [0]{\href }%
\providecommand \doibase [0]{http://dx.doi.org/}%
\providecommand \selectlanguage [0]{\@gobble}%
\providecommand \bibinfo  [0]{\@secondoftwo}%
\providecommand \bibfield  [0]{\@secondoftwo}%
\providecommand \translation [1]{[#1]}%
\providecommand \BibitemOpen [0]{}%
\providecommand \bibitemStop [0]{}%
\providecommand \bibitemNoStop [0]{.\EOS\space}%
\providecommand \EOS [0]{\spacefactor3000\relax}%
\providecommand \BibitemShut  [1]{\csname bibitem#1\endcsname}%
\let\auto@bib@innerbib\@empty
\bibitem [{\citenamefont {Car}\ and\ \citenamefont
  {Parrinello}(1985)}]{car1985unified}%
  \BibitemOpen
  \bibfield  {author} {\bibinfo {author} {\bibfnamefont {R.}~\bibnamefont
  {Car}}\ and\ \bibinfo {author} {\bibfnamefont {M.}~\bibnamefont
  {Parrinello}},\ }\href@noop {} {\bibfield  {journal} {\bibinfo  {journal}
  {Physical Review Letters}\ }\textbf {\bibinfo {volume} {55}},\ \bibinfo
  {pages} {2471} (\bibinfo {year} {1985})}\BibitemShut {NoStop}%
\bibitem [{\citenamefont {Marx}\ and\ \citenamefont
  {Hutter}(2009)}]{marx2009ab}%
  \BibitemOpen
  \bibfield  {author} {\bibinfo {author} {\bibfnamefont {D.}~\bibnamefont
  {Marx}}\ and\ \bibinfo {author} {\bibfnamefont {J.}~\bibnamefont {Hutter}},\
  }\href@noop {} {\emph {\bibinfo {title} {Ab initio molecular dynamics: basic
  theory and advanced methods}}}\ (\bibinfo  {publisher} {Cambridge University
  Press},\ \bibinfo {year} {2009})\BibitemShut {NoStop}%
\bibitem [{\citenamefont {Kohn}\ and\ \citenamefont
  {Sham}(1965)}]{kohn1965self}%
  \BibitemOpen
  \bibfield  {author} {\bibinfo {author} {\bibfnamefont {W.}~\bibnamefont
  {Kohn}}\ and\ \bibinfo {author} {\bibfnamefont {L.~J.}\ \bibnamefont
  {Sham}},\ }\href@noop {} {\bibfield  {journal} {\bibinfo  {journal} {Physical
  Review}\ }\textbf {\bibinfo {volume} {140}},\ \bibinfo {pages} {A1133}
  (\bibinfo {year} {1965})}\BibitemShut {NoStop}%
\bibitem [{\citenamefont {Vanommeslaeghe}\ \emph {et~al.}(2010)\citenamefont
  {Vanommeslaeghe}, \citenamefont {Hatcher}, \citenamefont {Acharya},
  \citenamefont {Kundu}, \citenamefont {Shim}, \citenamefont {Darian},
  \citenamefont {Guvench}, \citenamefont {Lopes}, \citenamefont {Vorobyov},\
  and\ \citenamefont {Mackerell~Jr.}}]{vanommeslaeghe2010charmm}%
  \BibitemOpen
  \bibfield  {author} {\bibinfo {author} {\bibfnamefont {K.}~\bibnamefont
  {Vanommeslaeghe}}, \bibinfo {author} {\bibfnamefont {E.}~\bibnamefont
  {Hatcher}}, \bibinfo {author} {\bibfnamefont {C.}~\bibnamefont {Acharya}},
  \bibinfo {author} {\bibfnamefont {S.}~\bibnamefont {Kundu}, \bibfnamefont
  {S.and~Zhong}}, \bibinfo {author} {\bibfnamefont {J.}~\bibnamefont {Shim}},
  \bibinfo {author} {\bibfnamefont {E.}~\bibnamefont {Darian}}, \bibinfo
  {author} {\bibfnamefont {O.}~\bibnamefont {Guvench}}, \bibinfo {author}
  {\bibfnamefont {P.}~\bibnamefont {Lopes}}, \bibinfo {author} {\bibfnamefont
  {I.}~\bibnamefont {Vorobyov}}, \ and\ \bibinfo {author} {\bibfnamefont
  {A.}~\bibnamefont {Mackerell~Jr.}},\ }\href@noop {} {\bibfield  {journal}
  {\bibinfo  {journal} {Journal of Computational Chemistry}\ }\textbf {\bibinfo
  {volume} {31}},\ \bibinfo {pages} {671} (\bibinfo {year} {2010})}\BibitemShut
  {NoStop}%
\bibitem [{\citenamefont {Jorgensen}\ \emph {et~al.}(1996)\citenamefont
  {Jorgensen}, \citenamefont {Maxwell},\ and\ \citenamefont
  {Tirado-Rives}}]{jorgensen1996development}%
  \BibitemOpen
  \bibfield  {author} {\bibinfo {author} {\bibfnamefont {W.}~\bibnamefont
  {Jorgensen}}, \bibinfo {author} {\bibfnamefont {D.}~\bibnamefont {Maxwell}},
  \ and\ \bibinfo {author} {\bibfnamefont {J.}~\bibnamefont {Tirado-Rives}},\
  }\href@noop {} {\bibfield  {journal} {\bibinfo  {journal} {Journal of the
  American Chemical Society}\ }\textbf {\bibinfo {volume} {118}},\ \bibinfo
  {pages} {11225} (\bibinfo {year} {1996})}\BibitemShut {NoStop}%
\bibitem [{\citenamefont {Wang}\ \emph {et~al.}(2004)\citenamefont {Wang},
  \citenamefont {Wolf}, \citenamefont {Caldwell}, \citenamefont {Kollman},\
  and\ \citenamefont {Case}}]{wang2004development}%
  \BibitemOpen
  \bibfield  {author} {\bibinfo {author} {\bibfnamefont {J.}~\bibnamefont
  {Wang}}, \bibinfo {author} {\bibfnamefont {R.~M.}\ \bibnamefont {Wolf}},
  \bibinfo {author} {\bibfnamefont {J.~W.}\ \bibnamefont {Caldwell}}, \bibinfo
  {author} {\bibfnamefont {P.~A.}\ \bibnamefont {Kollman}}, \ and\ \bibinfo
  {author} {\bibfnamefont {D.~A.}\ \bibnamefont {Case}},\ }\href@noop {}
  {\bibfield  {journal} {\bibinfo  {journal} {Journal of Computational
  Chemistry}\ }\textbf {\bibinfo {volume} {25}},\ \bibinfo {pages} {1157}
  (\bibinfo {year} {2004})}\BibitemShut {NoStop}%
\bibitem [{\citenamefont {Stillinger}\ and\ \citenamefont
  {Weber}(1985)}]{stillinger1985SW}%
  \BibitemOpen
  \bibfield  {author} {\bibinfo {author} {\bibfnamefont {F.~H.}\ \bibnamefont
  {Stillinger}}\ and\ \bibinfo {author} {\bibfnamefont {T.~A.}\ \bibnamefont
  {Weber}},\ }\href@noop {} {\bibfield  {journal} {\bibinfo  {journal}
  {Physical Review B}\ }\textbf {\bibinfo {volume} {31}},\ \bibinfo {pages}
  {5262} (\bibinfo {year} {1985})}\BibitemShut {NoStop}%
\bibitem [{\citenamefont {Daw}\ and\ \citenamefont
  {Baskes}(1984)}]{daw1984EAM}%
  \BibitemOpen
  \bibfield  {author} {\bibinfo {author} {\bibfnamefont {M.~S.}\ \bibnamefont
  {Daw}}\ and\ \bibinfo {author} {\bibfnamefont {M.~I.}\ \bibnamefont
  {Baskes}},\ }\href@noop {} {\bibfield  {journal} {\bibinfo  {journal}
  {Physical Review B}\ }\textbf {\bibinfo {volume} {29}},\ \bibinfo {pages}
  {6443} (\bibinfo {year} {1984})}\BibitemShut {NoStop}%
\bibitem [{\citenamefont {Brenner}\ \emph {et~al.}(2002)\citenamefont
  {Brenner}, \citenamefont {Shenderova}, \citenamefont {Harrison},
  \citenamefont {Stuart}, \citenamefont {Ni},\ and\ \citenamefont
  {Sinnott}}]{brenner2002BOP}%
  \BibitemOpen
  \bibfield  {author} {\bibinfo {author} {\bibfnamefont {D.~W.}\ \bibnamefont
  {Brenner}}, \bibinfo {author} {\bibfnamefont {O.~A.}\ \bibnamefont
  {Shenderova}}, \bibinfo {author} {\bibfnamefont {J.~A.}\ \bibnamefont
  {Harrison}}, \bibinfo {author} {\bibfnamefont {S.~J.}\ \bibnamefont
  {Stuart}}, \bibinfo {author} {\bibfnamefont {B.}~\bibnamefont {Ni}}, \ and\
  \bibinfo {author} {\bibfnamefont {S.~B.}\ \bibnamefont {Sinnott}},\
  }\href@noop {} {\bibfield  {journal} {\bibinfo  {journal} {Journal of
  Physics: Condensed Matter}\ }\textbf {\bibinfo {volume} {14}},\ \bibinfo
  {pages} {783} (\bibinfo {year} {2002})}\BibitemShut {NoStop}%
\bibitem [{\citenamefont {Van~Duin}\ \emph {et~al.}(2001)\citenamefont
  {Van~Duin}, \citenamefont {Dasgupta}, \citenamefont {Lorant},\ and\
  \citenamefont {Goddard}}]{van2001rff}%
  \BibitemOpen
  \bibfield  {author} {\bibinfo {author} {\bibfnamefont {A.~C.}\ \bibnamefont
  {Van~Duin}}, \bibinfo {author} {\bibfnamefont {S.}~\bibnamefont {Dasgupta}},
  \bibinfo {author} {\bibfnamefont {F.}~\bibnamefont {Lorant}}, \ and\ \bibinfo
  {author} {\bibfnamefont {W.~A.}\ \bibnamefont {Goddard}},\ }\href@noop {}
  {\bibfield  {journal} {\bibinfo  {journal} {The Journal of Physical Chemistry
  A}\ }\textbf {\bibinfo {volume} {105}},\ \bibinfo {pages} {9396} (\bibinfo
  {year} {2001})}\BibitemShut {NoStop}%
\bibitem [{\citenamefont {Thompson}\ \emph {et~al.}(2015)\citenamefont
  {Thompson}, \citenamefont {Swiler}, \citenamefont {Trott}, \citenamefont
  {Foiles},\ and\ \citenamefont {Tucker}}]{thompson2015SNAP}%
  \BibitemOpen
  \bibfield  {author} {\bibinfo {author} {\bibfnamefont {A.~P.}\ \bibnamefont
  {Thompson}}, \bibinfo {author} {\bibfnamefont {L.~P.}\ \bibnamefont
  {Swiler}}, \bibinfo {author} {\bibfnamefont {C.~R.}\ \bibnamefont {Trott}},
  \bibinfo {author} {\bibfnamefont {S.~M.}\ \bibnamefont {Foiles}}, \ and\
  \bibinfo {author} {\bibfnamefont {G.~J.}\ \bibnamefont {Tucker}},\
  }\href@noop {} {\bibfield  {journal} {\bibinfo  {journal} {Journal of
  Computational Physics}\ }\textbf {\bibinfo {volume} {285}},\ \bibinfo {pages}
  {316} (\bibinfo {year} {2015})}\BibitemShut {NoStop}%
\bibitem [{\citenamefont {Huan}\ \emph {et~al.}(2017)\citenamefont {Huan},
  \citenamefont {Batra}, \citenamefont {Chapman}, \citenamefont {Krishnan},
  \citenamefont {Chen},\ and\ \citenamefont {Ramprasad}}]{huan2017AGNI}%
  \BibitemOpen
  \bibfield  {author} {\bibinfo {author} {\bibfnamefont {T.~D.}\ \bibnamefont
  {Huan}}, \bibinfo {author} {\bibfnamefont {R.}~\bibnamefont {Batra}},
  \bibinfo {author} {\bibfnamefont {J.}~\bibnamefont {Chapman}}, \bibinfo
  {author} {\bibfnamefont {S.}~\bibnamefont {Krishnan}}, \bibinfo {author}
  {\bibfnamefont {L.}~\bibnamefont {Chen}}, \ and\ \bibinfo {author}
  {\bibfnamefont {R.}~\bibnamefont {Ramprasad}},\ }\href@noop {} {\bibfield
  {journal} {\bibinfo  {journal} {NPJ Computational Materials}\ }\textbf
  {\bibinfo {volume} {3}},\ \bibinfo {pages} {1} (\bibinfo {year}
  {2017})}\BibitemShut {NoStop}%
\bibitem [{\citenamefont {Behler}\ and\ \citenamefont
  {Parrinello}(2007)}]{behler2007generalized}%
  \BibitemOpen
  \bibfield  {author} {\bibinfo {author} {\bibfnamefont {J.}~\bibnamefont
  {Behler}}\ and\ \bibinfo {author} {\bibfnamefont {M.}~\bibnamefont
  {Parrinello}},\ }\href@noop {} {\bibfield  {journal} {\bibinfo  {journal}
  {Physical Review Letters}\ }\textbf {\bibinfo {volume} {98}},\ \bibinfo
  {pages} {146401} (\bibinfo {year} {2007})}\BibitemShut {NoStop}%
\bibitem [{\citenamefont {Behler}(2016)}]{behler2016perspective}%
  \BibitemOpen
  \bibfield  {author} {\bibinfo {author} {\bibfnamefont {J.}~\bibnamefont
  {Behler}},\ }\href@noop {} {\bibfield  {journal} {\bibinfo  {journal} {The
  Journal of Chemical Physics}\ }\textbf {\bibinfo {volume} {145}},\ \bibinfo
  {pages} {170901} (\bibinfo {year} {2016})}\BibitemShut {NoStop}%
\bibitem [{\citenamefont {Morawietz}\ \emph {et~al.}(2016)\citenamefont
  {Morawietz}, \citenamefont {Singraber}, \citenamefont {Dellago},\ and\
  \citenamefont {Behler}}]{morawietz2016van}%
  \BibitemOpen
  \bibfield  {author} {\bibinfo {author} {\bibfnamefont {T.}~\bibnamefont
  {Morawietz}}, \bibinfo {author} {\bibfnamefont {A.}~\bibnamefont
  {Singraber}}, \bibinfo {author} {\bibfnamefont {C.}~\bibnamefont {Dellago}},
  \ and\ \bibinfo {author} {\bibfnamefont {J.}~\bibnamefont {Behler}},\
  }\href@noop {} {\bibfield  {journal} {\bibinfo  {journal} {Proceedings of the
  National Academy of Sciences}\ ,\ \bibinfo {pages} {201602375}} (\bibinfo
  {year} {2016})}\BibitemShut {NoStop}%
\bibitem [{\citenamefont {Bart{\'o}k}\ \emph {et~al.}(2010)\citenamefont
  {Bart{\'o}k}, \citenamefont {Payne}, \citenamefont {Kondor},\ and\
  \citenamefont {Cs{\'a}nyi}}]{bartok2010gaussian}%
  \BibitemOpen
  \bibfield  {author} {\bibinfo {author} {\bibfnamefont {A.~P.}\ \bibnamefont
  {Bart{\'o}k}}, \bibinfo {author} {\bibfnamefont {M.~C.}\ \bibnamefont
  {Payne}}, \bibinfo {author} {\bibfnamefont {R.}~\bibnamefont {Kondor}}, \
  and\ \bibinfo {author} {\bibfnamefont {G.}~\bibnamefont {Cs{\'a}nyi}},\
  }\href@noop {} {\bibfield  {journal} {\bibinfo  {journal} {Physical Review
  Letters}\ }\textbf {\bibinfo {volume} {104}},\ \bibinfo {pages} {136403}
  (\bibinfo {year} {2010})}\BibitemShut {NoStop}%
\bibitem [{\citenamefont {Rupp}\ \emph {et~al.}(2012)\citenamefont {Rupp},
  \citenamefont {Tkatchenko}, \citenamefont {M{\"u}ller},\ and\ \citenamefont
  {VonLilienfeld}}]{rupp2012fast}%
  \BibitemOpen
  \bibfield  {author} {\bibinfo {author} {\bibfnamefont {M.}~\bibnamefont
  {Rupp}}, \bibinfo {author} {\bibfnamefont {A.}~\bibnamefont {Tkatchenko}},
  \bibinfo {author} {\bibfnamefont {K.-R.}\ \bibnamefont {M{\"u}ller}}, \ and\
  \bibinfo {author} {\bibfnamefont {O.~A.}\ \bibnamefont {VonLilienfeld}},\
  }\href@noop {} {\bibfield  {journal} {\bibinfo  {journal} {Physical Review
  Letters}\ }\textbf {\bibinfo {volume} {108}},\ \bibinfo {pages} {058301}
  (\bibinfo {year} {2012})}\BibitemShut {NoStop}%
\bibitem [{\citenamefont {Sch{\"u}tt}\ \emph {et~al.}(2017)\citenamefont
  {Sch{\"u}tt}, \citenamefont {Arbabzadah}, \citenamefont {Chmiela},
  \citenamefont {M{\"u}ller},\ and\ \citenamefont
  {Tkatchenko}}]{schutt2017quantum}%
  \BibitemOpen
  \bibfield  {author} {\bibinfo {author} {\bibfnamefont {K.~T.}\ \bibnamefont
  {Sch{\"u}tt}}, \bibinfo {author} {\bibfnamefont {F.}~\bibnamefont
  {Arbabzadah}}, \bibinfo {author} {\bibfnamefont {S.}~\bibnamefont {Chmiela}},
  \bibinfo {author} {\bibfnamefont {K.~R.}\ \bibnamefont {M{\"u}ller}}, \ and\
  \bibinfo {author} {\bibfnamefont {A.}~\bibnamefont {Tkatchenko}},\
  }\href@noop {} {\bibfield  {journal} {\bibinfo  {journal} {Nature
  Communications}\ }\textbf {\bibinfo {volume} {8}},\ \bibinfo {pages} {13890}
  (\bibinfo {year} {2017})}\BibitemShut {NoStop}%
\bibitem [{\citenamefont {Chmiela}\ \emph {et~al.}(2017)\citenamefont
  {Chmiela}, \citenamefont {Tkatchenko}, \citenamefont {Sauceda}, \citenamefont
  {Poltavsky}, \citenamefont {Sch{\"u}tt},\ and\ \citenamefont
  {M{\"u}ller}}]{chmiela2017machine}%
  \BibitemOpen
  \bibfield  {author} {\bibinfo {author} {\bibfnamefont {S.}~\bibnamefont
  {Chmiela}}, \bibinfo {author} {\bibfnamefont {A.}~\bibnamefont {Tkatchenko}},
  \bibinfo {author} {\bibfnamefont {H.~E.}\ \bibnamefont {Sauceda}}, \bibinfo
  {author} {\bibfnamefont {I.}~\bibnamefont {Poltavsky}}, \bibinfo {author}
  {\bibfnamefont {K.~T.}\ \bibnamefont {Sch{\"u}tt}}, \ and\ \bibinfo {author}
  {\bibfnamefont {K.-R.}\ \bibnamefont {M{\"u}ller}},\ }\href@noop {}
  {\bibfield  {journal} {\bibinfo  {journal} {Science Advances}\ }\textbf
  {\bibinfo {volume} {3}},\ \bibinfo {pages} {e1603015} (\bibinfo {year}
  {2017})}\BibitemShut {NoStop}%
\bibitem [{\citenamefont {Smith}\ \emph {et~al.}(2017)\citenamefont {Smith},
  \citenamefont {Isayev},\ and\ \citenamefont {Roitberg}}]{smith2017ani}%
  \BibitemOpen
  \bibfield  {author} {\bibinfo {author} {\bibfnamefont {J.~S.}\ \bibnamefont
  {Smith}}, \bibinfo {author} {\bibfnamefont {O.}~\bibnamefont {Isayev}}, \
  and\ \bibinfo {author} {\bibfnamefont {A.~E.}\ \bibnamefont {Roitberg}},\
  }\href@noop {} {\bibfield  {journal} {\bibinfo  {journal} {Chemical Science}\
  }\textbf {\bibinfo {volume} {8}},\ \bibinfo {pages} {3192} (\bibinfo {year}
  {2017})}\BibitemShut {NoStop}%
\bibitem [{\citenamefont {Han}\ \emph {et~al.}(2018)\citenamefont {Han},
  \citenamefont {Zhang}, \citenamefont {Car},\ and\ \citenamefont
  {E}}]{han2017deep}%
  \BibitemOpen
  \bibfield  {author} {\bibinfo {author} {\bibfnamefont {J.}~\bibnamefont
  {Han}}, \bibinfo {author} {\bibfnamefont {L.}~\bibnamefont {Zhang}}, \bibinfo
  {author} {\bibfnamefont {R.}~\bibnamefont {Car}}, \ and\ \bibinfo {author}
  {\bibfnamefont {W.}~\bibnamefont {E}},\ }\href@noop {} {\bibfield  {journal}
  {\bibinfo  {journal} {Communications in Computational Physics}\ }\textbf
  {\bibinfo {volume} {23}},\ \bibinfo {pages} {629} (\bibinfo {year}
  {2018})}\BibitemShut {NoStop}%
\bibitem [{\citenamefont {Goodfellow}\ \emph {et~al.}(2016)\citenamefont
  {Goodfellow}, \citenamefont {Bengio},\ and\ \citenamefont
  {Courville}}]{goodfellow2016dl}%
  \BibitemOpen
  \bibfield  {author} {\bibinfo {author} {\bibfnamefont {I.}~\bibnamefont
  {Goodfellow}}, \bibinfo {author} {\bibfnamefont {Y.}~\bibnamefont {Bengio}},
  \ and\ \bibinfo {author} {\bibfnamefont {A.}~\bibnamefont {Courville}},\
  }\href@noop {} {\emph {\bibinfo {title} {Deep learning}}}\ (\bibinfo
  {publisher} {MIT Press},\ \bibinfo {year} {2016})\BibitemShut {NoStop}%
\bibitem [{\citenamefont {Bengio}\ \emph {et~al.}(2007)\citenamefont {Bengio},
  \citenamefont {Lamblin}, \citenamefont {Popovici},\ and\ \citenamefont
  {Larochelle}}]{bengio2007greedy}%
  \BibitemOpen
  \bibfield  {author} {\bibinfo {author} {\bibfnamefont {Y.}~\bibnamefont
  {Bengio}}, \bibinfo {author} {\bibfnamefont {P.}~\bibnamefont {Lamblin}},
  \bibinfo {author} {\bibfnamefont {D.}~\bibnamefont {Popovici}}, \ and\
  \bibinfo {author} {\bibfnamefont {H.}~\bibnamefont {Larochelle}},\ }in\
  \href@noop {} {\emph {\bibinfo {booktitle} {Advances in neural information
  processing systems}}}\ (\bibinfo {year} {2007})\ pp.\ \bibinfo {pages}
  {153--160}\BibitemShut {NoStop}%
\bibitem [{\citenamefont {Krizhevsky}\ \emph {et~al.}(2012)\citenamefont
  {Krizhevsky}, \citenamefont {Sutskever},\ and\ \citenamefont
  {Hinton}}]{krizhevsky2012imagenet}%
  \BibitemOpen
  \bibfield  {author} {\bibinfo {author} {\bibfnamefont {A.}~\bibnamefont
  {Krizhevsky}}, \bibinfo {author} {\bibfnamefont {I.}~\bibnamefont
  {Sutskever}}, \ and\ \bibinfo {author} {\bibfnamefont {G.~E.}\ \bibnamefont
  {Hinton}},\ }in\ \href@noop {} {\emph {\bibinfo {booktitle} {Advances in
  neural information processing systems}}}\ (\bibinfo {year} {2012})\ pp.\
  \bibinfo {pages} {1097--1105}\BibitemShut {NoStop}%
\bibitem [{\citenamefont {Kingma}\ and\ \citenamefont
  {Ba}(2015)}]{Kingma2015adam}%
  \BibitemOpen
  \bibfield  {author} {\bibinfo {author} {\bibfnamefont {D.}~\bibnamefont
  {Kingma}}\ and\ \bibinfo {author} {\bibfnamefont {J.}~\bibnamefont {Ba}},\
  }in\ \href@noop {} {\emph {\bibinfo {booktitle} {Proceedings of the
  International Conference on Learning Representations (ICLR)}}}\ (\bibinfo
  {year} {2015})\BibitemShut {NoStop}%
\bibitem [{\citenamefont {Adamo}\ and\ \citenamefont
  {Barone}(1999)}]{Carlo1999PBE0}%
  \BibitemOpen
  \bibfield  {author} {\bibinfo {author} {\bibfnamefont {C.}~\bibnamefont
  {Adamo}}\ and\ \bibinfo {author} {\bibfnamefont {V.}~\bibnamefont {Barone}},\
  }\href@noop {} {\bibfield  {journal} {\bibinfo  {journal} {The Journal of
  Chemical Physics}\ }\textbf {\bibinfo {volume} {110}},\ \bibinfo {pages}
  {6158} (\bibinfo {year} {1999})}\BibitemShut {NoStop}%
\bibitem [{\citenamefont {Tkatchenko}\ and\ \citenamefont
  {Scheffler}(2009)}]{TS2009TS}%
  \BibitemOpen
  \bibfield  {author} {\bibinfo {author} {\bibfnamefont {A.}~\bibnamefont
  {Tkatchenko}}\ and\ \bibinfo {author} {\bibfnamefont {M.}~\bibnamefont
  {Scheffler}},\ }\href@noop {} {\bibfield  {journal} {\bibinfo  {journal}
  {Physical Review Letters}\ }\textbf {\bibinfo {volume} {102}},\ \bibinfo
  {pages} {073005} (\bibinfo {year} {2009})}\BibitemShut {NoStop}%
\bibitem [{\citenamefont {Perdew}\ \emph {et~al.}(1996)\citenamefont {Perdew},
  \citenamefont {Burke},\ and\ \citenamefont {Ernzerhof}}]{Perdew1996PBE}%
  \BibitemOpen
  \bibfield  {author} {\bibinfo {author} {\bibfnamefont {J.~P.}\ \bibnamefont
  {Perdew}}, \bibinfo {author} {\bibfnamefont {K.}~\bibnamefont {Burke}}, \
  and\ \bibinfo {author} {\bibfnamefont {M.}~\bibnamefont {Ernzerhof}},\
  }\href@noop {} {\bibfield  {journal} {\bibinfo  {journal} {Physical Review
  Letters}\ }\textbf {\bibinfo {volume} {77}},\ \bibinfo {pages} {3865}
  (\bibinfo {year} {1996})}\BibitemShut {NoStop}%
\bibitem [{\citenamefont {Ceriotti}\ \emph {et~al.}(2014)\citenamefont
  {Ceriotti}, \citenamefont {More},\ and\ \citenamefont
  {Manolopoulos}}]{Ceriotti2014iPI}%
  \BibitemOpen
  \bibfield  {author} {\bibinfo {author} {\bibfnamefont {M.}~\bibnamefont
  {Ceriotti}}, \bibinfo {author} {\bibfnamefont {J.}~\bibnamefont {More}}, \
  and\ \bibinfo {author} {\bibfnamefont {D.~E.}\ \bibnamefont {Manolopoulos}},\
  }\href@noop {} {\bibfield  {journal} {\bibinfo  {journal} {Computer Physics
  Communications}\ }\textbf {\bibinfo {volume} {185}},\ \bibinfo {pages} {1019}
  (\bibinfo {year} {2014})}\BibitemShut {NoStop}%
\bibitem [{\citenamefont {Jorgensen}\ \emph {et~al.}(1983)\citenamefont
  {Jorgensen}, \citenamefont {Chandrasekhar}, \citenamefont {Madura},
  \citenamefont {Impey},\ and\ \citenamefont {Klein}}]{Jorgensen1983TIP3P}%
  \BibitemOpen
  \bibfield  {author} {\bibinfo {author} {\bibfnamefont {W.~L.}\ \bibnamefont
  {Jorgensen}}, \bibinfo {author} {\bibfnamefont {J.}~\bibnamefont
  {Chandrasekhar}}, \bibinfo {author} {\bibfnamefont {J.~D.}\ \bibnamefont
  {Madura}}, \bibinfo {author} {\bibfnamefont {R.~W.}\ \bibnamefont {Impey}}, \
  and\ \bibinfo {author} {\bibfnamefont {M.~L.}\ \bibnamefont {Klein}},\
  }\href@noop {} {\bibfield  {journal} {\bibinfo  {journal} {The Journal of
  Chemical Physics}\ }\textbf {\bibinfo {volume} {79}},\ \bibinfo {pages} {926}
  (\bibinfo {year} {1983})}\BibitemShut {NoStop}%
\bibitem [{\citenamefont {Pronk}\ \emph {et~al.}(2013)\citenamefont {Pronk},
  \citenamefont {P{\'a}ll}, \citenamefont {Schulz}, \citenamefont {Larsson},
  \citenamefont {Bjelkmar}, \citenamefont {Apostolov}, \citenamefont {Shirts},
  \citenamefont {Smith}, \citenamefont {Kasson}, \citenamefont {van~der Spoel},
  \citenamefont {Hess},\ and\ \citenamefont {Lindahl}}]{pronk2013gromacs}%
  \BibitemOpen
  \bibfield  {author} {\bibinfo {author} {\bibfnamefont {S.}~\bibnamefont
  {Pronk}}, \bibinfo {author} {\bibfnamefont {S.}~\bibnamefont {P{\'a}ll}},
  \bibinfo {author} {\bibfnamefont {R.}~\bibnamefont {Schulz}}, \bibinfo
  {author} {\bibfnamefont {P.}~\bibnamefont {Larsson}}, \bibinfo {author}
  {\bibfnamefont {P.}~\bibnamefont {Bjelkmar}}, \bibinfo {author}
  {\bibfnamefont {R.}~\bibnamefont {Apostolov}}, \bibinfo {author}
  {\bibfnamefont {M.}~\bibnamefont {Shirts}}, \bibinfo {author} {\bibfnamefont
  {J.}~\bibnamefont {Smith}}, \bibinfo {author} {\bibfnamefont
  {P.}~\bibnamefont {Kasson}}, \bibinfo {author} {\bibfnamefont
  {D.}~\bibnamefont {van~der Spoel}}, \bibinfo {author} {\bibfnamefont
  {B.}~\bibnamefont {Hess}}, \ and\ \bibinfo {author} {\bibfnamefont
  {E.}~\bibnamefont {Lindahl}},\ }\href@noop {} {\bibfield  {journal} {\bibinfo
   {journal} {Bioinformatics}\ ,\ \bibinfo {pages} {btt055}} (\bibinfo {year}
  {2013})}\BibitemShut {NoStop}%
\bibitem [{\citenamefont {Andreussi}\ \emph {et~al.}(2017)\citenamefont
  {Andreussi}, \citenamefont {Brumme}, \citenamefont {Bunau}, \citenamefont
  {Nardelli}, \citenamefont {Calandra}, \citenamefont {Car}, \citenamefont
  {Cavazzoni}, \citenamefont {Ceresoli}, \citenamefont {Cococcioni},
  \citenamefont {Colonna} \emph {et~al.}}]{QE-2017}%
  \BibitemOpen
  \bibfield  {author} {\bibinfo {author} {\bibfnamefont {O.}~\bibnamefont
  {Andreussi}}, \bibinfo {author} {\bibfnamefont {T.}~\bibnamefont {Brumme}},
  \bibinfo {author} {\bibfnamefont {O.}~\bibnamefont {Bunau}}, \bibinfo
  {author} {\bibfnamefont {M.~B.}\ \bibnamefont {Nardelli}}, \bibinfo {author}
  {\bibfnamefont {M.}~\bibnamefont {Calandra}}, \bibinfo {author}
  {\bibfnamefont {R.}~\bibnamefont {Car}}, \bibinfo {author} {\bibfnamefont
  {C.}~\bibnamefont {Cavazzoni}}, \bibinfo {author} {\bibfnamefont
  {D.}~\bibnamefont {Ceresoli}}, \bibinfo {author} {\bibfnamefont
  {M.}~\bibnamefont {Cococcioni}}, \bibinfo {author} {\bibfnamefont
  {N.}~\bibnamefont {Colonna}},  \emph {et~al.},\ }\href@noop {} {\bibfield
  {journal} {\bibinfo  {journal} {Journal of Physics: Condensed Matter}\ }
  (\bibinfo {year} {2017})}\BibitemShut {NoStop}%
\bibitem [{\citenamefont {Watts}\ \emph {et~al.}(1993)\citenamefont {Watts},
  \citenamefont {Gauss},\ and\ \citenamefont {Bartlett}}]{Watts1993CCSDT}%
  \BibitemOpen
  \bibfield  {author} {\bibinfo {author} {\bibfnamefont {J.~D.}\ \bibnamefont
  {Watts}}, \bibinfo {author} {\bibfnamefont {J.}~\bibnamefont {Gauss}}, \ and\
  \bibinfo {author} {\bibfnamefont {R.~J.}\ \bibnamefont {Bartlett}},\
  }\href@noop {} {\bibfield  {journal} {\bibinfo  {journal} {The Journal of
  Chemical Physics}\ }\textbf {\bibinfo {volume} {98}},\ \bibinfo {pages}
  {8718} (\bibinfo {year} {1993})}\BibitemShut {NoStop}%
\bibitem [{\citenamefont {Ceperley}\ and\ \citenamefont
  {Alder}(1986)}]{ceperley1986qmc}%
  \BibitemOpen
  \bibfield  {author} {\bibinfo {author} {\bibfnamefont {D.}~\bibnamefont
  {Ceperley}}\ and\ \bibinfo {author} {\bibfnamefont {B.}~\bibnamefont
  {Alder}},\ }\href@noop {} {\bibfield  {journal} {\bibinfo  {journal}
  {Science}\ }\textbf {\bibinfo {volume} {231}} (\bibinfo {year}
  {1986})}\BibitemShut {NoStop}%
\end{thebibliography}%




\begin{thebibliography}{8}%
\makeatletter
\providecommand \@ifxundefined [1]{%
 \@ifx{#1\undefined}
}%
\providecommand \@ifnum [1]{%
 \ifnum #1\expandafter \@firstoftwo
 \else \expandafter \@secondoftwo
 \fi
}%
\providecommand \@ifx [1]{%
 \ifx #1\expandafter \@firstoftwo
 \else \expandafter \@secondoftwo
 \fi
}%
\providecommand \natexlab [1]{#1}%
\providecommand \enquote  [1]{``#1''}%
\providecommand \bibnamefont  [1]{#1}%
\providecommand \bibfnamefont [1]{#1}%
\providecommand \citenamefont [1]{#1}%
\providecommand \href@noop [0]{\@secondoftwo}%
\providecommand \href [0]{\begingroup \@sanitize@url \@href}%
\providecommand \@href[1]{\@@startlink{#1}\@@href}%
\providecommand \@@href[1]{\endgroup#1\@@endlink}%
\providecommand \@sanitize@url [0]{\catcode `\\12\catcode `\$12\catcode
  `\&12\catcode `\#12\catcode `\^12\catcode `\_12\catcode `\%12\relax}%
\providecommand \@@startlink[1]{}%
\providecommand \@@endlink[0]{}%
\providecommand \url  [0]{\begingroup\@sanitize@url \@url }%
\providecommand \@url [1]{\endgroup\@href {#1}{\urlprefix }}%
\providecommand \urlprefix  [0]{URL }%
\providecommand \Eprint [0]{\href }%
\providecommand \doibase [0]{http://dx.doi.org/}%
\providecommand \selectlanguage [0]{\@gobble}%
\providecommand \bibinfo  [0]{\@secondoftwo}%
\providecommand \bibfield  [0]{\@secondoftwo}%
\providecommand \translation [1]{[#1]}%
\providecommand \BibitemOpen [0]{}%
\providecommand \bibitemStop [0]{}%
\providecommand \bibitemNoStop [0]{.\EOS\space}%
\providecommand \EOS [0]{\spacefactor3000\relax}%
\providecommand \BibitemShut  [1]{\csname bibitem#1\endcsname}%
\let\auto@bib@innerbib\@empty
\bibitem [{\citenamefont {Ceriotti}, \citenamefont {Manolopoulos},\ and\
  \citenamefont {Parrinello}(2011)}]{Ceriotti2011GLE}%
  \BibitemOpen
  \bibfield  {author} {\bibinfo {author} {\bibnamefont {Ceriotti},
  \bibfnamefont {M.}}, \bibinfo {author} {\bibnamefont {Manolopoulos},
  \bibfnamefont {D.~E.}}, \ and\ \bibinfo {author} {\bibnamefont {Parrinello},
  \bibfnamefont {M.}},\ }\href@noop {} {\bibfield  {journal} {\bibinfo
  {journal} {The Journal of Chemical Physics}\ }\textbf {\bibinfo {volume}
  {134}},\ \bibinfo {pages} {084104} (\bibinfo {year} {2011})}\BibitemShut
  {NoStop}%
\bibitem [{\citenamefont {Ceriotti}, \citenamefont {More},\ and\ \citenamefont
  {Manolopoulos}(2014)}]{Ceriotti2014iPI}%
  \BibitemOpen
  \bibfield  {author} {\bibinfo {author} {\bibnamefont {Ceriotti},
  \bibfnamefont {M.}}, \bibinfo {author} {\bibnamefont {More}, \bibfnamefont
  {J.}}, \ and\ \bibinfo {author} {\bibnamefont {Manolopoulos}, \bibfnamefont
  {D.~E.}},\ }\href@noop {} {\bibfield  {journal} {\bibinfo  {journal}
  {Computer Physics Communications}\ }\textbf {\bibinfo {volume} {185}},\
  \bibinfo {pages} {1019} (\bibinfo {year} {2014})}\BibitemShut {NoStop}%
\bibitem [{\citenamefont {Han}\ \emph {et~al.}(2017)\citenamefont {Han},
  \citenamefont {Zhang}, \citenamefont {Car},\ and\ \citenamefont
  {E}}]{han2017deep}%
  \BibitemOpen
  \bibfield  {author} {\bibinfo {author} {\bibnamefont {Han}, \bibfnamefont
  {J.}}, \bibinfo {author} {\bibnamefont {Zhang}, \bibfnamefont {L.}}, \bibinfo
  {author} {\bibnamefont {Car}, \bibfnamefont {R.}}, \ and\ \bibinfo {author}
  {\bibnamefont {E}, \bibfnamefont {W.}},\ }\href@noop {} {\bibfield  {journal}
  {\bibinfo  {journal} {arXiv Preprint arXiv:1707.01478}\ } (\bibinfo {year}
  {2017})}\BibitemShut {NoStop}%
\bibitem [{\citenamefont {Ioffe}\ and\ \citenamefont
  {Szegedy}(2015)}]{Ioffe2015BN}%
  \BibitemOpen
  \bibfield  {author} {\bibinfo {author} {\bibnamefont {Ioffe}, \bibfnamefont
  {S.}}\ and\ \bibinfo {author} {\bibnamefont {Szegedy}, \bibfnamefont {C.}},\
  }in\ \href@noop {} {\emph {\bibinfo {booktitle} {Proceedings of The 32nd
  International Conference on Machine Learning (ICML)}}}\ (\bibinfo {year}
  {2015})\BibitemShut {NoStop}%
\bibitem [{\citenamefont {Kingma}\ and\ \citenamefont
  {Ba}(2015)}]{Kingma2015adam}%
  \BibitemOpen
  \bibfield  {author} {\bibinfo {author} {\bibnamefont {Kingma}, \bibfnamefont
  {D.}}\ and\ \bibinfo {author} {\bibnamefont {Ba}, \bibfnamefont {J.}},\ }in\
  \href@noop {} {\emph {\bibinfo {booktitle} {Proceedings of the International
  Conference on Learning Representations (ICLR)}}}\ (\bibinfo {year}
  {2015})\BibitemShut {NoStop}%
\bibitem [{\citenamefont {Lechner}\ and\ \citenamefont
  {Dellago}(2008)}]{lechner2008accurate}%
  \BibitemOpen
  \bibfield  {author} {\bibinfo {author} {\bibnamefont {Lechner}, \bibfnamefont
  {W.}}\ and\ \bibinfo {author} {\bibnamefont {Dellago}, \bibfnamefont {C.}},\
  }\href@noop {} {\bibfield  {journal} {\bibinfo  {journal} {The Journal of
  chemical physics}\ }\textbf {\bibinfo {volume} {129}},\ \bibinfo {pages}
  {114707} (\bibinfo {year} {2008})}\BibitemShut {NoStop}%
\bibitem [{\citenamefont {Martyna}, \citenamefont {Klein},\ and\ \citenamefont
  {Tuckerman}(1992)}]{Tuckerman1992NH}%
  \BibitemOpen
  \bibfield  {author} {\bibinfo {author} {\bibnamefont {Martyna}, \bibfnamefont
  {G.~J.}}, \bibinfo {author} {\bibnamefont {Klein}, \bibfnamefont {M.~L.}}, \
  and\ \bibinfo {author} {\bibnamefont {Tuckerman}, \bibfnamefont {M.}},\
  }\href@noop {} {\bibfield  {journal} {\bibinfo  {journal} {The Journal of
  Chemical Physics}\ }\textbf {\bibinfo {volume} {97}},\ \bibinfo {pages}
  {2635} (\bibinfo {year} {1992})}\BibitemShut {NoStop}%
\bibitem [{\citenamefont {Parrinello}\ and\ \citenamefont
  {Rahman}(1980)}]{Parrinello1980PR}%
  \BibitemOpen
  \bibfield  {author} {\bibinfo {author} {\bibnamefont {Parrinello},
  \bibfnamefont {M.}}\ and\ \bibinfo {author} {\bibnamefont {Rahman},
  \bibfnamefont {A.}},\ }\href@noop {} {\bibfield  {journal} {\bibinfo
  {journal} {Physical Review Letters}\ }\textbf {\bibinfo {volume} {45}},\
  \bibinfo {pages} {1196} (\bibinfo {year} {1980})}\BibitemShut {NoStop}%
\end{thebibliography}
%

{
\appendix
\section{Supplementary Materials}
\subsection{Training/Testing Data}
\subsubsection{water and ice}
The data used for training and/or testing are extracted from the AIMD
simulations summarized in Tab.~\ref{tab:thermo-cond}.
All the simulations adopt a time step of 0.48 fs.
The PI-AIMD simulations use the CPMD codes of Quantum Espresso
\footnote{http://www.quantum-espresso.org/} for the DFT part
and are interfaced with the i-PI code \cite{Ceriotti2014iPI} 
for the path-integral part.
The generalized Langevin equation with color noise \cite{Ceriotti2011GLE}
in i-PI requires 8 beads 
for a converged representation of the Feynman paths.
The classical AIMD simulations use the CPMD codes of Quantum Espresso
and adopt the Nos\'e-Hoover thermostat \cite{Tuckerman1992NH} 
for thermalization.
The Parrinello-Rahman technique~\cite{Parrinello1980PR} 
for variable cell dynamics is adopted in all cases. 
The training datasets include 
95000 snapshots (from 105000 total snapshots)
randomly selected along the liquid water trajectory,
19500 snapshots (from 24000 total snapshots)
randomly selected along the ice (b) trajectory,
9500 snapshots (from 12000 total snapshots)
randomly selected along the ice (c) trajectory,
and 9500 snapshots (from 12000 total snapshots)
randomly selected along the ice (d) trajectory.
The remaining snapshots in the database are used for testing purposes.

\begin{table}
\centering
\caption{Equilibrated AIMD trajectories (traj.) 
for liquid water (LW) and ice Ih.} 
\label{tab:thermo-cond}
\begin{tabular*}{0.8\textwidth}{@{\extracolsep{\fill}}
cccccc}\hline\hline
System  & PI/classical   & $N$&$P$ [bar]&$T$ [K] & traj. length [ps]   \\
\hline
LW      & path integral  & 64 & 1.0    & 300   & 6.2                 \\
ice (b) & path integral  & 96 & 1.0    & 273   & 1.5                 \\
ice (c) & classical      & 96 & 1.0    & 330   & 6.0                 \\
ice (d) & classical      & 96 & 2.13k  & 238   & 6.0                    \\\hline\hline
\end{tabular*}
\end{table}

\subsubsection{molecules}
The data and their complete description for the organic molecules
(benzene, uracil, napthalene, aspirin,
salicylic acid, malonaldehyde, ethanol, and toluene)
can be found at 
\href{http://quantum-machine.org/}{http://quantum-machine.org/}.
For each molecule, 95000 snapshots, randomly selected from the database,
are used to train the DeePMD model.
The remaining snapshots in the database are used for testing purposes.

\subsection{Implementation of the method}
The TensorFlow r1.0 software library 
~(\href{http://tensorflow.org/}{http://tensorflow.org/})
is interfaced with our C++ codes
for data training and
for calculating the energy, the forces, and the virial.

\subsubsection{network input data}
We consider a system consisting of $N$ atoms.
The global coordinates of the atoms, in the laboratory frame,
are $\left\{\bm{R}_1, \bm{R}_2, \dots, \bm{R}_N\right\}$, 
where $\bm{R}_i=\left\{x_i,y_i,z_i\right\}$ for each $i$.
The neighbors of atom $i$ are denoted by 
$\mn(i) = \left\{ j : \vert \bm{R}_{ij} \vert < R_c\right\}$,
where $\bm{R}_{ij} = \bm{R}_i - \bm{R}_j$, 
and $R_c$ is the cut-off radius.
The neighbor list $\mn(i)$ is sorted according to the scheme illustrated in Fig. 1.
In extended systems, 
the number of neighbors at different snapshots inside $R_c$ fluctuates.
Let $N_c$ be the largest fluctuating number of neighbors.
The two atoms used to define the axes of the local frame of atom $i$ 
are called the axis-atoms and 
are denoted by $a(i) \in \mn(i)$ and  $b(i) \in \mn(i)$, respectively. 
In general we choose two closest atoms, 
independently of their species, together with the center atom, to define the local frame.
Thus, in all the water cases, 
we choose the other two atoms belonging to the same water molecule.
We apply the same rule to the organic molecules, 
but in this case we exclude the hydrogen atoms 
in the definition of the axis-atoms.

Next, we define the rotation matrix 
$\mathcal{R}(\bm{R}_{ia(i)},\bm{R}_{ib(i)}) $ 
for the local frame of atom $i$,
\begin{align}
 \mathcal{R}(\bm{R}_{ia(i)},\bm{R}_{ib(i)}) =
 \begin{pmatrix}\bm{e}[\bm{R}_{ia(i)}]\\\bm{e}[\bm{R}_{ib(i)}-
(\bm{R}_{ia(i)}\cdot\bm{R}_{ib(i)}\bm{R}_{ia(i)}]\\
\bm{e}[\bm{R}_{ia(i)}\times\bm{R}_{ib(i)}]\end{pmatrix}^T,
\end{align}
where $\bm{e}[\bm{x}]\equiv\frac{\bm{x}}{||\bm{x}||}$. 
In this local frame of reference, 
we obtain the new set of coordinates:
\begin{align}
  \bm{R}_{ij}'=\left\{x_{ij}',y_{ij}',z_{ij}'\right\}
  =\left\{x_{ij},y_{ij},
  z_{ij}\right\}\mathcal{R}(\bm{R}_{ia(i)},\bm{R}_{ib(i)}),
\end{align}
and we define $R_{ij}' = ||\bm{R}_{ij}'||$. 
Then the spacial information for $j \in \mn(i)$ is 
\begin{align*}
  \bm{D}_{ij}\equiv
  \begin{cases}
  \left\{D_{ij}^0,D_{ij}^1,D_{ij}^2,
  D_{ij}^3\right\}
  =
  \left\{\frac{1}{R_{ij}'},
  \frac{x_{ij}'}{R_{ij}'^2},\frac{y_{ij}'}{R_{ij}'^2},
  \frac{z_{ij}'}{R_{ij}'^2}\right\},
  &\text{full radial and angular information;}\\
  \left\{D_{ij}^0\right\}
  =
  \left\{\frac{1}{R_{ij}'}\right\},
  &\text{radial information only.}
  \end{cases}
\end{align*}
When $\alpha=0,1,2,3$, 
full (radial plus angular) information is provided.
When $\alpha=0$, only radial information is used.
Note that for $j \in \mn(i)$, $D_{ij}^\alpha$ is a function of 
the global coordinates of three or four atoms:
\begin{align*}
  D_{ij}^\alpha=
 \begin{cases}
   D_{ij}^\alpha(\bm{R}_{i}, \bm{R}_{a(i)},
   \bm{R}_{b(i)}),&\text{for $j=a(i)$ or $j=b(i)$};\\
   D_{ij}^\alpha(\bm{R}_{i}, \bm{R}_{a(i)}, \bm{R}_{b(i)},
   \bm{R}_{j}),&\text{otherwise}.
\end{cases}
\end{align*}
This formula is useful 
in the derivation of the formulae for the forces and 
the virial tensor given below.

The neural network uses a fixed input data size.
Thus, when the size of $\mathcal{N}(i)$ is smaller than $N_c$, 
we temporarily set to zero the input nodes 
not used for storing the $D_{ij}^\alpha$.
The nodes set to zero are still labeled by $D_{ij}^\alpha$.

The $D_{ij}^\alpha$ 
are then standardized to be the input data for the neural networks.
In this procedure, 
the $D_{ij}^\alpha$ are grouped according to 
the different atomic species.
Within each group we calculate the mean and standard deviation
of each $D_{ij}^\alpha$
by averaging over the snapshots of the training sample
and over all the atoms in the group.
Then we shift the $D_{ij}^\alpha$ by their corresponding means,
and divide them by their corresponding standard deviations.
Because of the weight $1/R$ in the $D_{ij}^\alpha$ 
and because the unoccupied nodes are set to zero,
some standard deviations are very small or even zero. 
This causes an ill-posed training process.
Therefore, after the shift operations,
we divide by 0.01~\AA${}^{-1}$ the $D_{ij}^\alpha$ 
with standard deviation smaller than 0.01~\AA${}^{-1}$.
For simplicity, we still use the same notation 
for the standardized $D_{ij}^\alpha$.

\subsubsection{deep neural network for the energy}
For atom $i$, 
the ``atomic energy'' is represented as
\begin{align}
  E_i  = N_{\bm{w}(i)}( \{D_{ij}^\alpha\}_{j\in\mn(i), \alpha} ),
\end{align}
where $N_{\bm{w}(i)}$ is the network 
that computes the atomic contribution to the total energy, 
and $\bm{w}(i)$ are the weights used to parametrize the network, 
which depend on the chemical species of atom $i$. 

In this work, $N_{\bm{w}(i)}$ is constructed as a feedforward network
in which data flows from the input layer as $\{D_{ij}^\alpha\}$, through multiple fully connected hidden layers, to the output layer as the atomic energy $E_{i}$.
More specifically, a feedforward neural network with $N_h$ hidden layers is a mapping
\begin{align}
\label{eqn:map}
  \mn_{i} (\{D_{ij}^\alpha\})
  = \ml^\out_{i} \circ \ml_{i}^{N_h} \circ \ml_{i}^{N_h-1} \circ \cdots \circ \ml_{i}^1 (\{D_{ij}^\alpha\}),
\end{align}
where the symbol ``$\circ$'' denotes function composition.
Here $ \ml_{i}^{p} $ is the mapping from layer $p-1$ to $p$, a composition of a linear transformation and a non-linear transformation
\begin{equation}\label{eqn:layer}
  \vect d_i^p = \ml_{i}^p (\vect d_i^{p-1}) = \varphi \big( \vect W_{i}^p  \vect d_i^{p-1}  + \vect b_{i}^p   \big).
\end{equation}
The $\vect d^p_i \in \mathbb R^{M_p}$ denote the values of neurons in layer $p$ and $M_p$  the  number of neurons. The weight matrix $\vect W_{i}^p \in \mathbb R^{M_p\times M_{p-1}}$ and bias vector $\vect b_{i}^p \in \mathbb R^{M_p}$
are free parameters of the linear transformation that are to be optimized.
The non-linear activation function $\varphi$ is in general a component-wise function,
and here it is taken to be the hyperbolic tangent, i.e.,
\begin{align}
  \varphi (d_1, d_2, \dots, d_M) = (\tanh(d_1), \tanh(d_2), \dots, \tanh(d_M)).
\end{align}
The output mapping $\ml^\out_{i}$ is a linear transformation,
\begin{align}
  \ml^\out_{i} (\vect d^{N_h}_i) =  \vect W_{i}^\out  \vect d^{N_h-1}  + b_{i}^\out,
\end{align}
with weight vector $ \vect W_{i}^\out\in \mathbb R^{1\times M_{N_h}}$  and bias $b_i^\out\in \mathbb R$ being free parameters to be optimized as well. On the whole, all the free parameters associated with atom $i$ are
\begin{equation}
\bm{w}(i) = \{\vect W_{i}^1, \vect b_{i}^1, \vect W_{i}^2, \vect b_{i}^2, \cdots, \vect W_{i}^{N_h}, \vect b_{i}^{N_h}, \vect W_{i}^\out, b_i^\out\}.
\end{equation}
It should be stressed that, to guarantee the permutational symmetry, atoms of the same species share the same parameters $\bm{w}$.

\subsubsection{forces and virial tensor}
The total potential energy is the sum of the $E_i$.
Thus the forces are
\begin{align*}
  \bm{F}_i
  & = -\nabla_{\bm{R}_i} E  = -\sum_j \nabla_{\bm{R}_i} E_j = 
  -\sum_j \sum_{k\in\mn(j)} \nabla_{\bm{R}_i} 
  N_{\bm{w}(j)}( \{D_{jk}^\alpha\}_{k\in\mn(j), \alpha} )  \\
  & = -\sum_j \sum_{k\in\mn(j)} \sum_\alpha
  \frac{\partial N_{\bm{w}(j)} }{\partial D^\alpha_{jk}}
  \nabla_{\bm{R}_i} D^\alpha_{jk} \\
  & = - \sum_{k\in\mn(i)} \sum_\alpha
  \frac{\partial N_{\bm{w}(i)} }{\partial D^\alpha_{ik}} 
  \nabla_{\bm{R}_i} D^\alpha_{ik}
  -\sum_{j\neq i}\sum_{k\in\mn(j)} \sum_\alpha
  \delta (i-a(j))\frac{\partial N_{\bm{w}(j)} }{\partial D^\alpha_{jk}}
  \nabla_{\bm{R}_i} D^\alpha_{jk} \\
  &\quad
  -\sum_{j\neq i}\sum_{k\in\mn(j)} \sum_\alpha
  \delta (i-b(j))\frac{\partial N_{\bm{w}(j)} }
  {\partial D^\alpha_{jk}} \nabla_{\bm{R}_i} D^\alpha_{jk}
  -\sum_{j\neq i} \sum_{k\in\tilde\mn(j)} \sum_\alpha
  \delta (i-k)\frac{\partial N_{\bm{w}(j)} }{\partial D^\alpha_{jk}}
  \nabla_{\bm{R}_i} D^\alpha_{jk} \\
  & =- \sum_{k\in\mn(i)} \sum_\alpha
  \frac{\partial E_i }{\partial D^\alpha_{ik}} 
  \nabla_{\bm{R}_i} D^\alpha_{jk}
  -\sum_{j\neq i}\sum_{k\in\mn(j)} \sum_\alpha
  \delta (i-a(j))\frac{\partial E_j }
  {\partial D^\alpha_{jk}} \nabla_{\bm{R}_i} D^\alpha_{jk} \\
  &\quad
  -\sum_{j\neq i}\sum_{k\in\mn(j)} \sum_\alpha
  \delta (i-b(j))\frac{\partial E_j }
  {\partial D^\alpha_{jk}}\nabla_{\bm{R}_i} D^\alpha_{jk}
  -\sum_{j\neq i} \sum_{k\in\tilde\mn(j)} \sum_\alpha
  \delta (i-k)\frac{\partial E_j }{\partial D^\alpha_{jk}} 
  \nabla_{\bm{R}_i} D^\alpha_{jk},
\end{align*}
where $\tilde\mn(j)=\mn(j)\backslash\{a(j),b(j)\}$. 

The virial tensor is defined as 
$\Xi_{\alpha\beta}=-\frac{1}{2}\sum_i R_{i\alpha} F_{i\beta}$,
where the indices $\alpha$ and $\beta$ 
indicate Cartesian components in the lab reference frame.
Due to the periodic boundary conditions, 
one cannot directly use the absolute coordinates 
$R_{i\alpha}$ to compute the virial tensor.
Rather, in the AIMD framework, the virial tensor is defined with 
an alternative but equivalent formula, i.e.,
\begin{align}
  \Xi_{\alpha\beta}=-\frac{1}{2}\sum_{\gamma}
  \frac{\partial{E}}{\partial{h_{\alpha\gamma}}}h_{\gamma\beta},
\end{align}
where $h$ is the cell tensor.
In our framework, due to the decomposition of the local energy $E_i$,
one computes the virial tensor by:
\begin{align}
  \Xi_{\alpha\beta}=
  -\frac{1}{2}\sum_{i,j}x_\alpha^{(i,j)}{f}_\beta^{(i,j)}.
\end{align}
$x_\alpha^{(i,j)}$ is the $\alpha$-th component 
of the vector oriented from the $i$-th to the $j$-th atom 
in the difference:
\begin{align}
  x_\alpha^{(i,j)}=x_\alpha^{(i)}-x_\alpha^{(j)}.
\end{align}
${f}_\beta^{(i,j)}$ is the $\beta$-th component of 
the negative gradient of $E_i$ w.r.t. $x_j$, i.e.,
\begin{align}
  {f}_\beta^{(i,j)}=-\frac{\partial{E_i}}{\partial_{x_j^\beta}}.
\end{align}

Together with the energy representation described above, 
all the quantities needed for training and MD simulations, although complicated, have been analytically defined. 
In particular, it is noted that the derivatives of the total energy with respect to the atomic positions, appearing in both the forces and the viral tensor,
are computed by the chain rule through the backpropagation algorithm, provided by TensorFlow.
To make it work, we additionally implement the computation of $\nabla_{\bm{R}_i} D^\alpha_{jk}$ in C++ and interface it with TensorFlow.

\subsubsection{Training Details}
During the training process, 
one minimizes the family of loss functions defined in the paper:
\begin{align}\label{eqn:loss}
   L(p_\epsilon, p_f, p_\xi) = 
  {p_\epsilon} \Delta \epsilon^2 + \frac{p_f}{3N} 
  \sum_i \vert \Delta\vect F_i\vert^2 + \frac{p_\xi}{9} ||\Delta \xi||^2.
\end{align}

The network weights are optimized with the Adam stochastic gradient
descent method~\cite{Kingma2015adam}. 
An initial learning rate $r_{l0}=0.001$ is used with 
the Adam parameters set to 
$\beta_1$=0.9, $\beta_2$=0.999, and $\epsilon$=1.0$\times{10}^{-8}$, 
which are the default settings in TensorFlow. 
The learning rate $r_{l}$ decays exponentially with the global step:
\begin{align}
 r_l=r_{l0}d_r^{-c_s/d_s},
\end{align}
where $d_r$, $c_s$, and $d_s$ are the decay rate, 
the global step, and the decay step, respectively. 
In this paper, the batch size is 4 
in all the training processes. 
The decay rate is 0.95. 
For liquid water, 
the training process undergoes 4000000 steps in total,
and the learning rate is updated every 20000 steps. 
For molecules, 
the training process undergoes 8000000 steps in total, 
and the learning rate is updated every 40000 steps. 

We remark that, for the prefactors, 
a proper linear evolution with the learning rate 
speeds up dramatically the training process. 
We define this process by: 
\begin{align}
 p = p_{limit}(1-\frac{ r_l}{ r_{l0}})+p_{start}(\frac{ r_l}{ r_{l0}}),
\end{align}
in which $p_{start}$ is the prefactor at the beginning of the training process,
and $p_{limit}$ is approximately the prefactor at the end.
We define $p_{start}$ for the energy, the forces, and the virial as
$p_{estart}$, $p_{fstart}$, and $p_{vstart}$, respectively.
Similarly, we define $p_{limit}$ for the energy, the forces, 
and the virial as
$p_{elimit}$, $p_{flimit}$, and $p_{vlimit}$, respectively.
In this paper, we use the following scheme: 
 \begin{align}
 \begin{cases}
 p_{estart}=1,&  p_{elimit}=400;\\
   p_{fstart}=1000,&  p_{flimit}=1,\\
 \end{cases}
\end{align}
for both water and the molecules, and 
 \begin{align}
  \begin{cases}
 p_{vstart}=1,  p_{vlimit}=400,&\text{ for liquid water and ice (b);}\\
p_{vstart}=0,  p_{vlimit}=0, &\text{ for ice (c) and (d) and the molecules.}\\
 \end{cases}
\end{align}

The above scheme is based on the following considerations. 
Each snapshot of the AIMD trajectories provides 
1 energy, $3N$ forces, and 6 independent virial tensor elements. 
The number of force components is much larger than 
the number of energy and virial tensor components. 
Therefore, matching the forces at the very beginning 
of the training process makes the training efficient. 
As the training proceeds, 
increasing the prefactors of the energy and the virial tensor 
allows us to achieve a well balanced training 
in which the energy, the forces, 
and the virial are mutually consistent. 

In the original Deep Potential paper~\cite{han2017deep}, 
only the energy was used to train the network, 
requiring in some cases the use of Batch Normalization techniques
\cite{Ioffe2015BN} to deal with issues of overfitting and
training efficiency.
Adding the forces and/or the virial tensor provides a strong 
regularization of the network and makes training significantly more efficient.
Thus Batch Normalization techniques are not necessary 
within the DeePMD framework.

\subsubsection{DeePMD details}
In the path-integral/classical $NPT$ simulations of liquid water and ice,
we integrate our codes with the i-PI software.
The DeePMD simulations are performed at the same thermodynamic conditions, 
and use the same temperature and pressure controls,
of the corresponding AIMD simulations.
All DeePMD trajectories for water and ice are 300 ps long 
and use the same time step of the AIMD simulations.

We use our own code to perform the constant temperature MD simulations
of the organic molecules.
In each DeePMD simulation the temperature is the same of that of 
the corresponding AIMD simulation. 
The time step and time length of the trajectories in these simulations 
are the same of those in the corresponding AIMD trajectories.

\subsection{Additional Results}
The radial distribution functions (RDFs) of ice Ih (b), (c) ,and (d) are reported 
in Figs.~\ref{fig:rdf-ice-b}, ~\ref{fig:rdf-ice-c}, and~\ref{fig:rdf-ice-d}, respectively.

\begin{figure}
  \centering
  \includegraphics[width=0.48\textwidth]{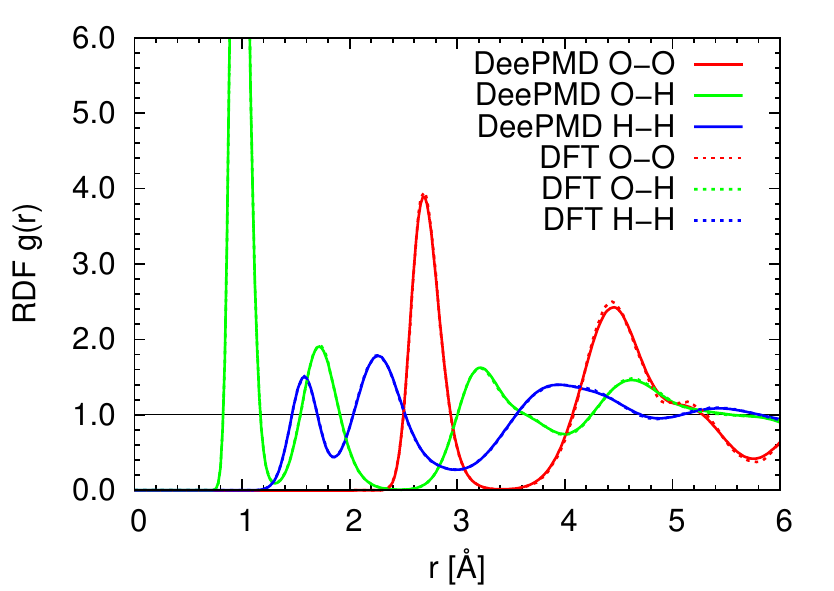}
  \caption{The comparison between the DeePMD RDFs 
  and the AIMD RDFs of ice Ih (b).
  }\label{fig:rdf-ice-b}
\end{figure}

\begin{figure}
  \centering
  \includegraphics[width=0.48\textwidth]{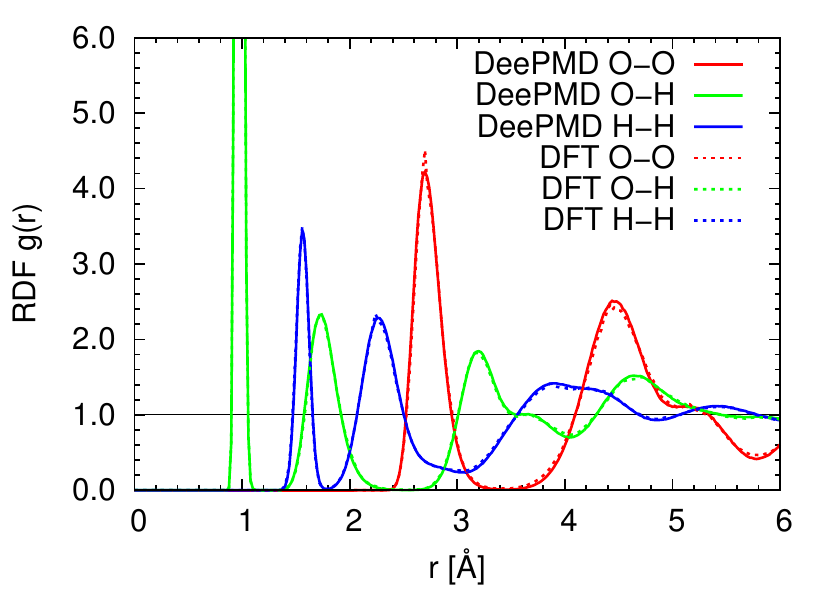}
  \caption{The comparison between the DeePMD RDFs 
  and the AIMD RDFs of ice Ih (c).
  }\label{fig:rdf-ice-c}
\end{figure}

\begin{figure}
  \centering
  \includegraphics[width=0.48\textwidth]{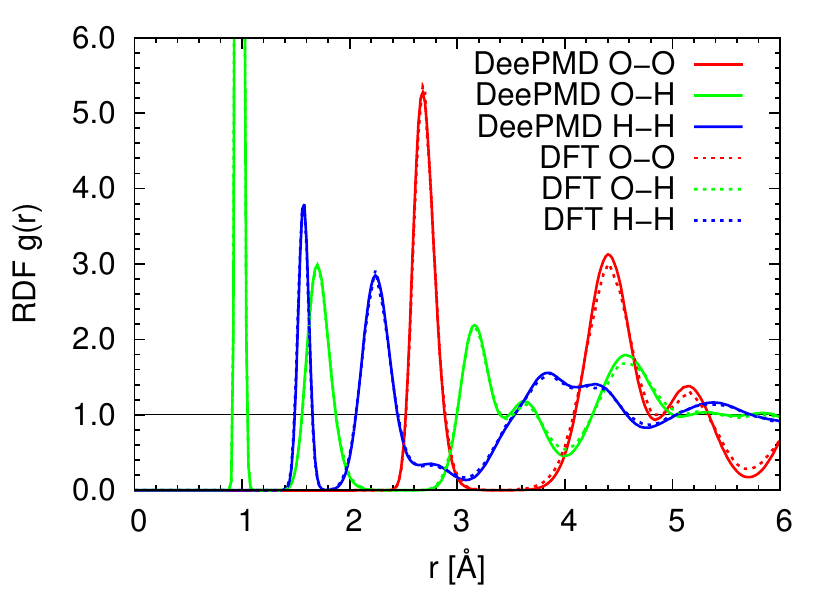}
  \caption{The comparison between the DeePMD RDFs 
  and the AIMD RDFs of ice Ih (d).
  }\label{fig:rdf-ice-d}
\end{figure}

The probability distribution function of the O-O bond orientation order parameter $Q_6$ is reported in Fig.~\ref{fig:q6}.
The bond orientation order parameter for oxygen $i$, as proposed in Ref.~\cite{lechner2008accurate}, is defined by 
\begin{align}
    Q_l(i) = 
  \Big[
  \frac{4\pi}{2l + 1}
  \sum_{m=-l}^l \vert \bar q_{lm}(i)\vert^2
  \Big]^{\frac12},
\end{align}
where
\begin{align}
  \bar q_{lm}(i) =
  \frac
  {\sum_{j\in \tilde N_b(i)} s(r_{ij}) q_{lm}(j)}
  {\sum_{j\in \tilde N_b(i)} s(r_{ij})}, ~~
   q_{lm}(i) =                                                                    
  \frac                                                                          
  {\sum_{j\in N_b(i)} s(r_{ij}) Y_{lm}(\hat {\vect r}_{ij})}               
  {\sum_{j\in N_b(i)} s(r_{ij})},
\end{align}
and $\tilde N_b(i) = N_b(i)\cup\{i\}$.
The $Y_{lm}(\cdots)$ denotes the spherical harmonic function,
the $N_b(i)$ denotes the set of oxygen neighbors of oxygen $i$, and 
the $s(r_{ij})$ is a switching function defined by
\begin{align}                                                                    
  s(r) =                                                                         
  \left \{                                                                       
  \begin{aligned}                                                                
    &1, & \quad  &r < r_{min}; \\                                                 
    &\frac12 + \frac12 \cos \Big(\pi \, \frac{r - r_{min}}{r_{max} - r_{min}} \Big), & \quad &r_{min}\leq r < r_{max};  \\                                        
    &0,  & \quad  &r \geq r_{max}.                                                
  \end{aligned}                                                                  
  \right.                                                                        
\end{align}
In this work we take $r_{min} = 0.31$~nm and $r_{max} = 0.36$~nm.

\begin{figure}
  \centering
  \includegraphics[width=0.48\textwidth]{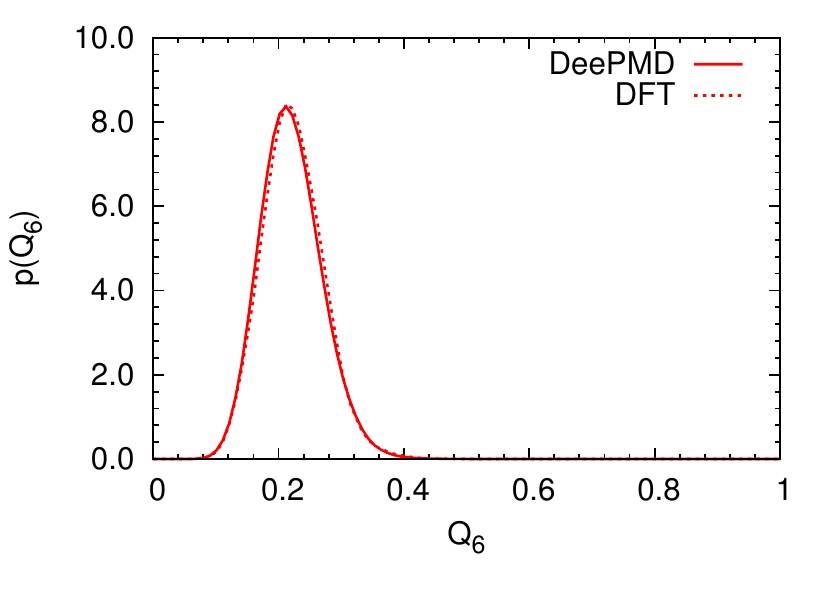}
  \caption{Probability distribution function of the O-O bond orientation order parameter $Q_6$}\label{fig:q6}
\end{figure}
\newpage
}

\end{document}